\documentclass[11pt,a4paper]{article}
\pdfoutput=1
\usepackage{jheppub}
\usepackage{amsmath}
\usepackage{dsfont}
\usepackage{MnSymbol}
\usepackage{array}
\usepackage{multirow}
\usepackage{hhline}
\usepackage[table,xcdraw]{xcolor}
\usepackage{ytableau}
\usepackage{tikz} 
\usepackage{epstopdf}

\newcommand{\assign}{:=}
\newcommand{\backassign}{=:}
\newcommand{\mathd}{\mathrm{d}}
\newcommand{\mathe}{\mathrm{e}}
\newcommand{\mathi}{\mathrm{i}}
\newcommand{\tautau}{\underline{\tau}_1,\underline{\tau}_2}
\newcommand{\lbracket}{\left\langle}
\newcommand{\rbracket}{\right\rangle}
\newcommand{\Z}{\mathcal{Z}}
\newcommand{\C}{\mathcal{C}}
\newcommand{\shift}{\hat{\mathsf{M}}_{i,\alpha}}
\newcommand\res[1]{\underset{#1}{\mathrm{Res}}}

\preprint{{\small{\textsf{UUITP-39/19}}}}

\title{Exact SUSY Wilson loops on $S^3$ from $q$-Virasoro constraints}

\author[a]{Luca Cassia,}
\author[a]{Rebecca Lodin,}
\author[a,b,c,d]{Aleksandr Popolitov,}
\author[a]{and Maxim Zabzine}

\affiliation[a]{Department of Physics and Astronomy, Uppsala University,\\
Box 516, SE-75120 Uppsala, Sweden.}
\affiliation[b]{Moscow Institute for Physics and Technology, Dolgoprudny, Russia}
\affiliation[c]{ITEP, Moscow 117218, Russia}
\affiliation[d]{Institute for Information Transmission Problems, Moscow 127994, Russia}

\emailAdd{luca.cassia@physics.uu.se}
\emailAdd{rebecca.lodin@physics.uu.se}
\emailAdd{popolit@gmail.com}
\emailAdd{maxim.zabzine@physics.uu.se}

\abstract{
  Using the ideas from the BPS/CFT correspondence,
  we give an explicit recursive formula for computing supersymmetric Wilson loop
  averages in 3d $\mathcal{N}=2$ Yang-Mills-Chern-Simons $U(N)$ theory on the
  squashed sphere $S^3_b$ with one adjoint chiral and two antichiral fundamental
  multiplets, for specific values of Chern-Simons level~$\kappa_2$ and Fayet-Illiopoulos
  parameter~$\kappa_1$. For these values of $\kappa_1$ and $\kappa_2$
  the north and south pole turn out to be completely independent, and therefore
  Wilson loop averages factorize into answers for the two constituent
  $D^2 \times S^1$ theories.
  In particular, our formula provides results for the theory on the round sphere
  when the squashing is removed.
}

\arxivnumber{1909.10352}

\begin{document}

\maketitle
\flushbottom

{\section{Introduction} \label{sec:introduction}

During the last 30 years there has been a vast development in  our understanding of supersymmetric gauge theories in various dimensions. For many supersymmetric gauge 
 theories the partition functions and the expectation values of certain protected BPS observables can be calculated exactly.  It has been observed that 
  these exact gauge theory quantities can be expressed in terms of two dimensional conformal field theories (or their deformations), a phenomenon known 
  as BPS/CFT correspondence \cite{Nekrasov-BPS/CFT-1, Nekrasov-BPS/CFT-2}.  
 One famous example of the BPS/CFT correspondence is  the AGT correspondence which relates the 4d $\mathcal{S}$-class theories to 2d Liouville and Toda models  \cite{Alday:2009aq,Wyllard:2009hg}. 
  By now we know many more concrete examples of the BPS/CFT correspondence.                 
   In this paper we concentrate on the concrete application of the BPS/CFT correspondence to 3d supersymmetric gauge theory. In particular, we will show how this correspondence 
  leads to explicit formulas for the expectation values of supersymmetric Wilson loops.  
  
Starting from the work \cite{Pestun:2007rz}, there has been many explicit calculations of the partition functions and other BPS observables for supersymmetric gauge theories on compact manifolds, 
 see \cite{Pestun:2016zxk} for a review. In these calculations the main tool is equivariant localization in the space of fields, and  
 the final answers are typically  expressed in terms of finite dimensional matrix models which are generically rather complicated. 
  In the case of the 3d ${\cal N}=2$  Yang-Mills-Chern-Simons (YM-CS) theories, the corresponding  matrix models were derived in
   \cite{Kapustin:2009kz} for the round sphere $S^3$ and in \cite{Hama:2011ea, Imamura:2011wg} for the squashed sphere $S^3_b$. The expectation value of the supersymmetric Wilson loop corresponds to the specific insertion 
    of a Schur polynomial in the matrix model and it is convenient to combine them into the generating function  $\Z \left(\tautau\right)$. For a generic value of the squashing parameter $b$, this generating function encodes 
     all Wilson loops in arbitrary representations. In \cite{Nedelin:2016gwu} it has been observed that  $\Z \left(\tautau\right)$ for 3d ${\cal N} = 2$ YM-CS $U(N)$ theory coupled
     to adjoint and possibly (anti)fundamental chiral multiplets has a free field representation in terms of vertex operators and screening charges of the $q$-Virasoro modular double
      (this observation is based  on earlier works \cite{Shiraishi:1995rp} and \cite{Aganagic:2013tta, Aganagic:2014kja}).
      Upon fixing some parameters, the generating function $\Z \left(\tautau\right)$ satisfies two commuting sets of $q$-Virasoro constraints which provide the Ward identities for the corresponding 
       matrix model. However, this free field construction is formal and it breaks down in the case of the round sphere $b=1$. 
     In this paper we would like to address  the numerous analytical issues and  solve these Ward identities explicitly.  
         
    The present paper is the development of the ideas and techniques of \cite{Lodin:2018lbz},
    where  the YM-CS living on $D^2 \times S^1$ was worked out in detail. 
        There the Ward identities (the $q$-Virasoro constraints) for the matrix model were derived by inserting certain $q$-difference operators under the 
        integral with some analytical issues being addressed, and finally the Ward identities lead to the iterative solution for all correlators in the corresponding 
        matrix model.
    Here  we consider a particular supersymmetric gauge theory
    -- the $\mathcal{N} = 2$ $U(N)$ Yang-Mills-Chern-Simons (YM-CS) theory  coupled to one adjoint and two fundamental anti-chiral matter multiplets
    on the
    squashed sphere background $S^3_b$
    (see Section~\ref{sec:gauge-theory-def} for the precise definition).
    Going to the squashed sphere case requires, among other things, a new careful analysis of the poles
    coming from contributions of gauge and matter multiplets
    (outlined in Appendix~\ref{sec:contour_shift}). This analysis needs to be performed
    case by case for every theory one considers. 
    How generic the class of theories for which our procedure works is therefore a subject of further research.
    At the end we provide a simple and efficient way to algorithmically calculate (supersymmetric) Wilson loop averages
    in any concrete representation. This procedure, which can be readily cast into computer program form,
    is the main new contribution of this paper.
 As a rough illustration of our result,  for a Wilson loop around the north ($\alpha = 1$) or south ($\alpha = 2$) pole
    in representation $\mathcal{R} = \{1,1\}$, i.e. the rank 2 antisymmetric, we get
    \begin{align}
      \langle \mathrm{WL}^{(\alpha)}_{\{1,1\}} \rangle =
      \frac{\left(t_\alpha-t_\alpha^{N}\right) \left(t_\alpha^{N}-1\right) \left(B (t_\alpha-1)-A^2 t_\alpha\right)}{B^2 (t_\alpha-1)^2 t_\alpha (t_\alpha+1)}~,
    \end{align}
    where $A$, $B$, and $t_\alpha$ are functions of the fundamental and adjoint masses and the squashing parameters
    (see Sections~\ref{sec:gauge-theory-def}~and~\ref{sec:derivation}). In this paper we will always be discussing {\it normalized} expectation values of Wilson loops. 

    Furthermore, even for the particular theory on $S^3_b$ we managed to get the whole
    scheme of \cite{Lodin:2018lbz} working only for special values of the Chern-Simons (CS) level~$\kappa_2$
    and Fayet-Illiopoulos (FI) parameter~$\kappa_1$ (see Section~\ref{sec:derivation} for details).
    While these restrictions on $\kappa_1$ and $\kappa_2$ appear to be purely technical,
    i.e. they are needed to drastically simplify parts of the computation,
    at the moment it is not clear how to lift them.
    Therefore, some conceptual underlying reason for these restrictions may exist.
    We hope to address and push these limitations in the future.
  
    In addition to their practical 3d gauge theory use, the present analysis and the corresponding  Ward identities
    \eqref{eq:combined_constraint} are interesting from a purely matrix model point of view
    as well. They are nothing but $q$-Virasoro constraints (see Section~\ref{sec:free_field}), where the choice
    of representation of the Heisenberg generators depends on the adjoint and antifundamental masses.
    This was anticipated already in \cite{Nedelin:2016gwu}, but in this paper we pay careful attention
    to various analytical issues, which required introduction of antifundamental multiplets. Thus, it puts the formal derivation of \cite{Nedelin:2016gwu} on a firm footing.
    Interestingly, as one takes the round sphere limit (Section~\ref{sec:limits}), the representation of the Heisenberg
    generators becomes singular and the free field representation fails. However, the Ward identities admit the well-defined limit and our result is still valid for round $S^3$.  
    It can be noted that the round sphere limit does not correspond to the standard semi-classical limit of the $q$-Virasoro algebra collapsing  
    to (usual) Virasoro algebra. 

    There is yet another angle from which our work may be interesting.
    Namely, a certain $q-$ (but not $t$) deformed matrix model (the BEM-model \cite{Brini:2011wi}) which calculates colored HOMFLY polynomials for torus knots.
    Generalizations of this model, both in the direction beyond torus knots and
    in the direction of turning on the $t$ parameter (i.e. going to Khovanov, Khovanov-Rozansky and superpolynomials)
    are much sought for. Once such $(q,t)$-deformation would be available, it would be very interesting
    to see how the techniques developed here, help to explain certain
    strange phenomena of the $(q,t)$-world such as chamber structures \cite{Dunin-Barkowski:2018kxg}
    and nimble evolution \cite{Anokhina:2019lbw}.
    
    The paper is organized as follows. We continue in Section~\ref{sec:gauge-theory-def} with the definition
    of the gauge theory that we consider. In Section~\ref{sec:derivation} we derive the $q$-Virasoro constraints using the insertion of a certain finite difference operator, and we also interpret the result using the free field representation. We then recursively solve the constraints and obtain explicit expressions for the expectation values of the first few supersymmetric Wilson loops in Section~\ref{sec:recursive_solution}. In Section~\ref{sec:limits} we take several interesting limits of the result. Finally we summarize and suggest directions for further research in Section~\ref{sec:conclusion}. Details of Schur polynomials and partitions, special functions and the difference operator and the shift of integration contour are left to the appendices.

\section{Gauge theory on the squashed 3-sphere}\label{sec:gauge-theory-def}
In this section we give the definition of the theory that we will be working with
and we also review some technical aspects regarding partition functions of 3d $\mathcal{N}=2$ YM-CS gauge theories.

We are interested in theories with unitary gauge group $U(N)$ and (anti-)chiral matter in the fundamental or adjoint representations.
Such theories can be placed on curved compact backgrounds while still preserving 2 supercharges as shown in
 \cite{Kapustin:2009kz, Hama:2011ea, Imamura:2011wg} and
 \cite{Festuccia:2011ws,Closset:2012ru,Closset:2013vra, Dumitrescu:2012ha,Closset:2014uda}. We work on a compact manifold of the form of a squashed 3-sphere $S^3_b$, defined as the locus
\begin{equation}\label{eq:squashing}
 \omega_1^2(x_1^2+x_2^2)+\omega_2^2(x_3^2+x_4^2)=1, \quad b^2 = \omega_2/\omega_1
\end{equation}
inside of $\mathbb{R}^4$, where $\omega_{1,2}$ are the (real) parameters of the squashing. It will also be useful to define the combination
\begin{equation}
\omega = \omega_1 + \omega_2~.
\end{equation}
Upon analytic continuation of the partition function we can take the squashing parameters to be arbitrary non-zero complex numbers.

Another useful way to describe the squashed sphere background is that of a singular elliptic fibration over an interval. In this picture, one of the two cycles of the torus fiber shrinks to a point at one edge of the base while the other cycle shrinks to a point at the opposite edge. If we cut open the interval at its midpoint, then the restriction of the fibration over each of the two smaller segments has the topology of a solid torus $D^2\times S^1$ with the degenerate fiber identified with the locus $\{0\}\times S^1$ (where $\{0\}$ is the center of the 2-disk). The gluing along the boundary $\partial(D^2\times S^1)\cong S^1\times S^1$ is done via a diffeomorphism that exchanges the two fundamental cycles of the torus.
Using the intrinsic coordinates $\theta,\phi,\chi$ given by
\begin{equation}
 \begin{split}
  x_1 & = {\omega_1}^{-1}\cos\theta\cos\phi ~, \\
  x_2 & = {\omega_1}^{-1}\cos\theta\sin\phi ~, \\
  x_3 & = {\omega_2}^{-1}\sin\theta\cos\chi ~, \\
  x_4 & = {\omega_2}^{-1}\sin\theta\sin\chi ~,
 \end{split}
\end{equation}
we can identify $\theta\in[0,\tfrac{\pi}{2}]$ with the coordinate along the base and $\phi,\chi\in[0,2\pi]$ with the coordinates on the torus fibers. The singular fibers are then given by the cycles at $\theta=0$ and $\theta=\tfrac{\pi}{2}$ which are parametrized by $\phi$ and $\chi$, respectively.

Upon using supersymmetric localization one finds that the 1-loop contributions of the gauge and matter multiplets are given in terms of products of double sine functions\footnote{The comparison between our notation and the literature is that $S_2 \left( \omega/2 -\mathi X | \underline{\omega} \right)=s_b\left(X\right)$, with $\omega_1 = \omega_2^{-1}=b$ and $\omega = Q$ \cite{Hama:2010av}.} $S_2(z|\underline{\omega})$ as summarized in Table~\ref{tab:1}.
\begin{table}[h]
\centering
\begin{tabular}{|l|l|}
\hline 
 multiplet & 1-loop contribution \\ 
\hline
 vector & $\prod_{\alpha\in\Delta} S_2\left(\alpha(\underline{X})|\underline{\omega}\right)$ \\
 chiral in irrep $\mathcal{R}$ & $\prod_{w\in\mathcal{R}} S_2\left(w(\underline{X})+M_{\mathcal{R}}|\underline{\omega}\right)^{-1}$ \\
 antichiral in irrep $\bar{\mathcal{R}}$ & $\prod_{w\in\mathcal{R}} S_2\left(-w(\underline{X})+M_{\bar{\mathcal{R}}}|\underline{\omega}\right)^{-1}$ \\ 
\hline
\end{tabular}
\caption{1-loop determinants of vector and (anti-)chiral multiplets.}
\label{tab:1}
\end{table}

Here $\alpha\in\Delta$ are the roots of the algebra (we always exclude the zero root) and $w$ are the weights of the representation $\mathcal{R}$,
while $M_\mathcal{R}$ are the masses of the (anti)chiral fields.
With $\underline{X}=(X_1,\dots,X_N)$ we indicate the \textit{gauge variables}, i.e. the integration variables of the localized partition function, taking values in the Cartan of the gauge group. For the case of $U(N)$ the roots are differences of fundamental weights $w_i$ so that we can write
\begin{equation}
 \alpha_{ij}(\underline{X}) = w_i(\underline{X})-w_j(\underline{X}) = X_i-X_j~,
\end{equation}
where the $X_i$ are imaginary numbers.

In addition to the 1-loop determinants we also allow for a CS term
\begin{equation}
 \prod_{i=1}^N\mathe^{-\frac{\pi\mathi\kappa_2}{\omega_1\omega_2}X_i^2}
\end{equation}
with level $\kappa_2\in\mathbb{Z}$ and, since the gauge group has a $U(1)$ in the center, an FI term
\begin{equation}
 \prod_{i=1}^N\mathe^{\frac{2\pi\mathi\kappa_1}{\omega_1\omega_2}X_i}
\end{equation}
with complexified parameter $\kappa_1\in\mathbb{C}$.

For technical reasons which will become clear in the following sections, we further specialize to a theory with a single $U(N)$ gauge group together with 1 adjoint massive chiral and 2 fundamental antichirals with masses $\mu,\nu\in\mathbb{C}$. The partition function can then be written explicitly as the integral\footnote{Up to an overall multiplicative factor, this partition function coincides with the ``level 6'' integral ${I\!I}^1_{n,(4,2)a}(\mu,\nu;-;\lambda;\tau)$ of \cite[Section 5.B]{VdB}, where the parameters are identified as
\[
  M_{\mathrm{a}} = \tau, \quad \kappa_1 = \lambda/2, \quad \kappa_2 = 1.
\]
Notice that the choice of CS level there is compatible with the one we have in Section~\ref{sec:derivation}.}
\begin{equation}\label{eq:partitionfunction}
\begin{split}
 \Z = \int_{(\mathi\mathbb{R})^N} \mathd X^N \;
 \underbrace{\prod_{k\neq j}\frac{S_2(X_k-X_j|\underline{\omega})}{S_2(X_k-X_j+M_{\mathrm{a}}|\underline{\omega})}}_{\Delta_S(\underline{X})}
 \prod_{i=1}^N & S_2(-X_i+\mu|\underline{\omega})^{-1} S_2(-X_i+\nu|\underline{\omega})^{-1} \\
 & \times \exp\left(-\frac{\pi\mathi\kappa_2}{\omega_1\omega_2}X_i^2 + \frac{2\pi\mathi\kappa_1}{\omega_1\omega_2}X_i \right)~.
\end{split}
\end{equation}
Following the mathematical literature we denote the combination of the vector's and adjoint chiral's 1-loop determinants as the function $\Delta_S(\underline{X})$ which from the point of view of the matrix model theory represents a generalization of the Vandermonde determinant of $U(N)$. For more details on this see \cite{VdB} and references therein.

Of great importance for the gauge theory is the computation of expectation values of supersymmetric Wilson loop operators. These correspond to quantum averages of traces of the holonomy of the gauge connection around some supersymmetric closed curves inside the spacetime manifold. Such supersymmetry preserving loops are referred to as $\tfrac{1}{2}$-BPS loops and, for generic $\omega_1,\omega_2$, are exactly the two singular fibers at $\theta=0$ and $\theta=\tfrac{\pi}{2}$.
Whenever the ratio of the squashing parameters is a rational number we also have a second family of $\tfrac{1}{2}$-BPS loops at $\theta\neq 0,\tfrac{\theta}{2}$ and wrapping the regular fibers according to the equation $\omega_1\phi+\omega_2\chi=const$ \cite{Tanaka:2012nr}. By definition these cycles are torus knots inside of $S^3_b$. In this paper we will only consider the case in which the Wilson loops wrap one or both of the singular fibers. 
Concretely, the traces are taken over arbitrary irreducible representations $\mathcal{R}_\rho$ of the gauge Lie algebra which for $U(N)$ are given as functions of the Cartan variables $X_i$ by the Schur polynomials
\begin{equation}
 \lbracket\mathrm{WL}_{\rho}^{(\alpha)} \rbracket = \lbracket\mathrm{Tr}_{\mathcal{R}_\rho}\left( \mathe^{\frac{2\pi\mathi}{\omega_\alpha}\underline{X}} \right)\rbracket
 = \lbracket s_\rho\left( \left\{\mathe^{\frac{2\pi\mathi}{\omega_\alpha}X_i} \right\}\right)\rbracket~,
\end{equation}
where the irreducible representations of the unitary group are labeled by Young diagrams, or equivalently, integer partitions $\rho$.
We provide more details on this in Appendix~\ref{sec:schur}.
Observe that the dependence of the Wilson loop on the label $\alpha$ tells us on which of the two supersymmetric cycles the holonomy is evaluated.

By introducing the auxiliary set of \textit{power sum} variables $p_s$ defined as
\begin{equation}
 p_s \left( \mathe^{\frac{2\pi\mathi}{\omega_\alpha}X_1} ,\dots, \mathe^{\frac{2\pi\mathi}{\omega_\alpha}X_N} \right)  = \sum_{i=1}^N \left(\mathe^{\frac{2\pi\mathi}{\omega_\alpha}X_i}\right)^s ~,
\end{equation}
i.e. the $s$-fold multiply-wound Wilson loop in the fundamental representation,
there is a canonical and algorithmic way to write the Schur polynomials as polynomial combinations of the $p_s$'s. This is a consequence of the well known fact that both $\{s_\rho\}$ and $\{p_s\}$ form a basis for the space of symmetric polynomials in the variables $\mathe^{\frac{2\pi\mathi}{\omega_\alpha}X_i}$, of which the Wilson loop operators are an example.
Then we can encode the expectation values of all such operators into a \textit{generating function}
\begin{equation} \label{eq:generating_function}
\begin{split}
  \Z \left(\tautau\right) = \int_{(\mathi\mathbb{R})^N} \mathd X^N \;
  & \Delta_S \left( \underline{X} \right) \prod_{i=1}^N S_2 \left( - X_i +
  \mu | \underline{\omega} \right)^{- 1} S_2 \left( - X_i +
  \nu | \underline{\omega} \right)^{- 1} \\
  & \times \exp \left( - \frac{\pi \mathi
  \kappa_2}{\omega_1 \omega_2} X_i^2 + \frac{2 \pi \mathi \kappa_1}{\omega_1
  \omega_2} X_i \right) \prod_{\alpha=1,2} \exp \left( \sum_{s = 1}^{\infty} \tau_{s, \alpha}
  \mathe^{\frac{2\pi\mathi s}{\omega_\alpha}X_i} \right)~,
  \end{split}
\end{equation}
where we introduced two infinite sets of formal \textit{time} variables $\tau_{s,\alpha}$ conjugate to the $p_s$, using the shorthand notation $\underline{\tau}_\alpha=\left(\tau_{1,\alpha} , \tau_{2,\alpha}, \dots\right)$. By taking derivatives in times of the generating function and subsequently setting all the times to zero we automatically get all expectation values of the power sum variables and consequently of the Schur polynomials, in other words the $\mathrm{WL}_{\rho}^{(\alpha)}$. Our goal in the following sections will be that of computing recursively such expectation values by making use of matrix models techniques.

For later convenience we also define the exponentiated variables\footnote{Another common parametrization used for instance in \cite{Nedelin:2016gwu} is that in which $q_\alpha$ and $t_\alpha$ are related to each other via $t_\alpha=(q_\alpha)^\beta$ for $\beta\in\mathbb{C}$. In this case $\beta$ can be naturally related to the parameter of the $\beta$-deformation of the Hermitian matrix model \cite{DiFrancesco:1993cyw,Morozov:1994hh,Odake:1999un,Awata:2009ur}.}
\begin{equation}\label{eq:exponentiated_variables}
\begin{array}{rcl}
 q_{\alpha} &=& \mathe^{\frac{2 \pi \mathi \omega}{\omega_{\alpha}}} \\
 t_{\alpha} &=& \mathe^{\frac{2 \pi \mathi  M_{\mathrm{a}}}{\omega_{\alpha}}} \\
 u_{\alpha} &=& \mathe^{\frac{2 \pi \mathi \mu}{\omega_{\alpha}}} \\
 v_{\alpha} &=& \mathe^{\frac{2 \pi \mathi \nu}{\omega_{\alpha}}} \\
 \lambda_{i,\alpha} &=& \mathe^{\frac{2 \pi \mathi X_i}{\omega_{\alpha}}}
\end{array}
\end{equation}
which, as remarked in \cite{Nedelin:2016gwu}, provide a natural way to describe the generating function as a vector in a representation of two commuting copies of the $q$-Virasoro algebra (see Section~\ref{sec:free_field} for more details on this). As a convenient notational shorthand we will also be using the variable $p_\alpha$ with $p_\alpha=q_\alpha t_\alpha^{-1}$ (not to be confused with the power sum variables $\{p_s\}$).

We remark here that for generic values of the squashing parameters $\omega_\alpha$ in the region where $\mathrm{Im}\left( \frac{\omega_1}{\omega_2}\right) \neq 0$ (see \eqref{eq:S2factorization}), the contribution of the fundamental antichiral multiplets can be written as 
\begin{equation} \label{eq:antichirals}
\begin{split}
\log\big[S_2 ( - X_i + \mu | \underline{\omega} )^{-1}
    & S_2 ( - X_i + \nu | \underline{\omega} )^{-1} \big] =  \\
     = & \frac{\mathi\pi}{\omega_1\omega_2}\left(X_i^2-X_i(\mu+\nu-\omega)+
\frac{\omega^2+\omega_1\omega_2}{6}+\frac{(\mu^2+\nu^2)-\omega(\mu+\nu)}{2} \right) \\
 & - \sum_{\alpha=1,2}\sum_{s=1}^\infty \frac{\mathe^{\frac{2\pi\mathi s}{\omega_\alpha}\omega}\left(\mathe^{-\frac{2\pi\mathi s}{\omega_\alpha}\mu}+\mathe^{-\frac{2\pi\mathi s}{\omega_\alpha}\nu}\right)\,\mathe^{\frac{2\pi\mathi s}{\omega_\alpha}X_i}}{s(1-\mathe^{\frac{2\pi\mathi s}{\omega_\alpha}\omega})}~.
\end{split}
\end{equation}
This contribution can equivalently be obtained (up to a numerical factor) via a redefinition of the CS level and FI parameter together with a shift of the time variables \cite{Nedelin:2016gwu}. The corresponding shifts are 
\begin{equation} \label{eq:shifts}
\begin{split}
 \kappa_1 \rightarrow \kappa_1 -\frac{\left( \mu + \nu - \omega \right)}{2}~,
 \quad \kappa_2 \rightarrow \kappa_2 - 1~,
 \quad \tau_{s,\alpha} \rightarrow \tau_{s,\alpha} - \frac{q_\alpha^s\left(u_\alpha^{-s}+v_\alpha^{-s}\right)}{s(1-q_\alpha^s)}~.
\end{split}
\end{equation}
However this transformation becomes singular in the round sphere limit as we discuss in Section~\ref{sec:limits}.

In what follows we will for brevity also use
\begin{equation}
 \label{eq:def1}
  \lbracket f \rbracket_\tau = \int_{(\mathi\mathbb{R})^N} \mathd X^N f\left(
  \underline{X} \right) J \left(
  \underline{X} | \tautau \right)~,
\end{equation}
where 
\begin{equation}\label{eq:J_definition}
\begin{split}
  J \left( \underline{X} | \tautau \right) = \Delta_S \left(
  \underline{X} \right) \prod_{i = 1}^N & S_2 \left( - X_i +
  \mu | \underline{\omega} \right)^{- 1} S_2 \left( - X_i +
  \nu | \underline{\omega} \right)^{- 1} \\
  & \times \exp \left( - \frac{\pi \mathi
  \kappa_2}{\omega_1 \omega_2} X_i^2 + \frac{2 \pi \mathi \kappa_1}{\omega_1
  \omega_2} X_i \right) \prod_{\alpha=1,2} \exp \left( \sum_{s = 1}^{\infty} \tau_{s, \alpha}
  \lambda_{i, \alpha}^s \right)
  \end{split}
\end{equation}
is the integrand of \eqref{eq:generating_function}, where it should be noted that $\lbracket f \rbracket_\tau$ still has a dependence on the times $\underline{\tau}_1$ and $\underline{\tau}_2$ (hence the label). With this notation we then have $\lbracket 1 \rbracket_\tau \equiv \Z \left( \tautau \right)$ and more generally
\begin{equation}
 \lbracket p_{s_1}(\left\{\lambda_{i,\alpha_1}\right\}) \dots p_{s_\ell}(\left\{\lambda_{i,\alpha_\ell}\right\}) \rbracket_\tau
 = \frac{\partial}{\partial\tau_{s_1,\alpha_1}} \dots \frac{\partial}{\partial\tau_{s_\ell,\alpha_\ell}} \Z(\tautau)~.
\end{equation}

In the next section we present the procedure to obtain the $q$-Virasoro constraints on the generating function $\Z\left(\tautau \right)$ using matrix model techniques.

\section{Derivation of $q$-Virasoro constraints} \label{sec:derivation}
For the gauge theory described above, we would now like to derive the $q$-Virasoro constraints using the trick of inserting a suitably chosen $q$-difference operator under the integral. This procedure will be very similar to that in \citep{Lodin:2018lbz}, the main difference being that now we are working in the logarithmic variables $X\sim\ln\lambda$.
\subsection{Definitions}
The finite difference operator $\shift$ we will use in order to derive the $q$-Virasoro constraints is defined as:
\begin{equation} \label{eq:shift_operator}
\begin{split}
  \hat{\mathsf{M}}_{i,1} f \left( \underline{X} \right) = & f (\ldots, X_i -
  \omega_{2}, \ldots) \\
  \hat{\mathsf{M}}_{i,2} f \left( \underline{X} \right) = & f (\ldots, X_i -
  \omega_{1}, \ldots)~,
  \end{split}
\end{equation}
where the notation is inspired by \cite{book:kc}.
In other words, $\shift$ corresponds to the operator that sends $\lambda_{i, \alpha}$ to
$\lambda_{i, \alpha} / q_{\alpha}$ (as can be seen from \eqref{eq:exponentiated_variables}). What we now wish to compute is the insertion of 
\begin{equation}\label{eq:insertion}
\sum_{i=1}^N \shift
\left[ \frac{1}{z
  \lambda_{i, \alpha}\left(1 - z
  \lambda_{i, \alpha}\right)} G_{i, \alpha} \left( \underline{\lambda} \right) \ldots \right]
\end{equation}
under the integral in \eqref{eq:generating_function} with $\dots$ denoting the integrand and
\begin{equation}\label{eq:G_function}
  G_{i, \alpha} \left( \underline{\lambda} \right) =   \prod_{j \neq i} \frac{1 - t_{\alpha} \lambda_{i,
  \alpha} / \lambda_{j, \alpha}}{1 - \lambda_{i, \alpha} / \lambda_{j,
  \alpha}} =   \prod_{j \neq i} \mathe^{\frac{\pi \mathi}{\omega_{\alpha}} M_{\mathrm{a}}} 
  \frac{\sin \left( \frac{\pi}{\omega_{\alpha}} \left( X_i - X_j +
  M_{\mathrm{a}} \right) \right)}{\sin \left( \frac{\pi}{\omega_{\alpha}} (X_i
  - X_j) \right)}~.
\end{equation}
For this reason we will be treating only one copy (i.e. one value of $\alpha=1,2$) at the time. The idea is then on the one hand to compute the action of the operator $\shift$ on the integrand, and on the other hand to trade this finite difference operator for a redefinition of the variables $X_i$ together with a shift of integration domain (where the details are outlined in Appendix~\ref{sec:contour_shift}). We then wish to equate these two computations and obtain the desired constraint.

The motivation for the form of this insertion can be seen as follows. The fraction appearing in \eqref{eq:insertion} can be written as 
\begin{equation}\label{eq:z-series}
\frac{1}{z\lambda_{i,\alpha}\left( 1-z\lambda_{i,\alpha}\right)} = \sum_{n = -1}^\infty \left(z\lambda_{i,\alpha}\right)^n ~,
\end{equation}
where the $z$-dependence is formal. This enables us to expand the obtained constraint in powers of $z$, generating a separate equation for each power (i.e. a set of Ward identities in the spirit of those of \cite{Nedelin:2015mio}).
Notice here that the summation in \eqref{eq:z-series} starts from $-1$ which in the language of \cite{Lodin:2018lbz} corresponds to considering generic and special constraints simultaneously.
This is in analogy with the derivation of the usual Virasoro constraints where one considers the insertion of the differential operator
$\tfrac{\partial}{\partial X_i}\left[X_i^{n+1}\dots\right]$ inside of the integral, only now we have to substitute the usual derivative with an appropriate $q$-difference operator, namely the combination $G_{i, \alpha}\left( \underline{\lambda} \right) \shift$.
Furthermore, the precise form of the function $G_{i, \alpha} \left( \underline{\lambda} \right)$ is necessary in order to introduce the desired pole structure (as shown in Appendix~\ref{sec:contour_shift}). We remark that its form is highly reminiscent of that of the Macdonald-Ruijsenaars operator $D_{q,t}$ \cite{Ruijsenaars1987,Shiraishi:1995rp} (although the exact relation is yet to be determined).

\subsection{Computing the insertion}\label{sec:computing_insertion}
As explained above, what we now wish to evaluate is the following equation
\begin{equation}\label{eq:starting_eq}
\begin{split}
  \left(\mathrm{LHS}\right){\assign} &
  \sum_{i=1}^{N}\int_{\hat{\mathsf{M}}^{-1}_{i,\alpha}C}{\mathd}X^{N}\left[{\frac{1}{z{\lambda}_{i,{\alpha}}\left(1-z
  {\lambda}_{i,{\alpha}}\right)}}G_{i,{\alpha}}\left({\underline{\lambda}}\right)J\left({\underline{X}}{|}{\tautau}\right)\right]=\\
  &
  \qquad \qquad =\int_{C}{\mathd}X^{N}\sum_{i=1}^{N}\shift \left[{\frac{1}{z{\lambda}_{i,{\alpha}}\left(1-z
  {\lambda}_{i,{\alpha}}\right)}}G_{i,{\alpha}}\left({\underline{\lambda}}\right)J\left({\underline{X}}{|}{\tautau}\right)\right]{\backassign}\left(\mathrm{RHS}\right)~,
\end{split}
\end{equation}
where the contour of integration $C$ is taken to be a middle dimensional subspace of $\mathbb{C}^N$ such that all the integration variables are purely imaginary, i.e. $C=(\mathi\mathbb{R})^N\subset\mathbb{C}^N$ (see Appendix~\ref{sec:contour_shift} for more details on this).

Here we have introduced the notation (LHS) for the left hand side of the equation and (RHS) for the right hand side so that we can discuss them separately. The (LHS) has been obtained by trading the finite difference operator with a redefinition of the integration variables $X_i$ and a shift of the integration domain, which as shown in Appendix~\ref{sec:contour_shift}, leaves the integral unchanged.
To then evaluate the (RHS) we will compute the action of $\shift$ on all the terms above. 
We remark that while \eqref{eq:starting_eq} holds for physical (real) values of the squashing parameters, all subsequent manipulations of this section are valid for any complex values of the parameters $\omega_1$ and $\omega_2$.

Starting with the (LHS), \eqref{eq:starting_eq} will hold at each order in $z$ separately, and so we can use the algebraic identity
\begin{equation}
  \sum_{i = 1}^N \frac{1}{z \lambda_{i, \alpha} (1 - z \lambda_{i, \alpha})}
  \underbrace{\left[ \prod_{j \neq i} \frac{t_{\alpha} \lambda_{i, \alpha} - \lambda_{j,
  \alpha}}{\lambda_{i, \alpha} - \lambda_{j, \alpha}} \right]}_{G_{i,\alpha}} =
  {\frac{1}{z} \sum_{i = 1}^N \frac{1}{\lambda_{i, \alpha}}}
  + {\frac{1}{1 - t_{\alpha}}} -
  {\frac{t_{\alpha}^N}{1 - t_{\alpha}}  \prod_{j = 1}^N
  \frac{1 - t^{- 1}_{\alpha} z \lambda_{j, \alpha}}{1 - z \lambda_{j, \alpha}}}
\end{equation}
to evaluate the (LHS). Thus
\begin{equation} \label{eq:LHS}
\begin{split}
  \left(\mathrm{LHS}\right) &
  =\lbracket
    {\frac{1}{z} \sum_{i = 1}^N \frac{1}{\lambda_{i, \alpha}}}
  + {\frac{1}{1 - t_{\alpha}}} -
  {\frac{t_{\alpha}^N}{1 - t_{\alpha}}  \prod_{j = 1}^N
  \frac{1 - t^{- 1}_{\alpha} z \lambda_{j, \alpha}}{1 - z \lambda_{j, \alpha}}}
  \rbracket_\tau\\
  &
  ={\frac{1}{z}}\sum_{i=1}^{N}\lbracket {\frac{1}{{\lambda}_{i,{\alpha}}}}\rbracket_\tau
   +{\frac{1}{1-t_{{\alpha}}}}\Z\left(\tautau\right)
   -{\frac{t_{{\alpha}}^{N}}{1-t_{{\alpha}}}} \exp\left(\sum_{s=1}^{{\infty}}z^{s}{\frac{\left(1-t_{{\alpha}}^{-s}\right)}{s}} \, {\frac{{\partial}}{{\partial}{\tau}_{s,{\alpha}}}}\right)\Z\left(\tautau\right)~,
\end{split}
\end{equation}
using the $\lbracket \dots \rbracket_\tau$ notation defined in \eqref{eq:def1}. Notice that the first term in the second line is an expectation value of a negative power of $\lambda_{i,\alpha}$ which we require to be canceled by a similar term on the RHS, as we do not have an interpretation for such terms as differential operators acting on the generating function.

Let us now proceed to the (RHS) of \eqref{eq:starting_eq} by computing the variations, in other words the action of $\shift$, on each of the terms in the insertion and the integrand of \eqref{eq:generating_function} separately. Starting with the insertion, the variation of $G_{i,{\alpha}}\left({\underline{\lambda}}\right)$ is 
\begin{equation} \label{eq:G_variation}
\begin{split}
  \shift G_{i,{\alpha}}\left({\underline{\lambda}}\right) = G_{i,{\alpha}}\left({\underline{\lambda}}\right)\prod_{j \neq i} \frac{1 - t_{\alpha} \lambda_{i,
  \alpha} q_\alpha^{-1}/ \lambda_{j, \alpha}}{1 - \lambda_{i, \alpha}q_\alpha^{-1} / \lambda_{j,
  \alpha}} \prod_{j \neq i} \frac{1 -  \lambda_{i,
  \alpha}/ \lambda_{j, \alpha}}{1 - t_{\alpha} \lambda_{i, \alpha} / \lambda_{j,
  \alpha}}~.
\end{split}
\end{equation}
We can then use the quasi-periodicity property in \eqref{eq:quasiperiodicity} to compute the variation of the double sine
\begin{equation}
  \shift S_2 \left( X_i | \underline{\omega} \right) =
  S_2 \left( X_i | \underline{\omega} \right) \left[- 2 \sin \left( \tfrac{\pi}{\omega_{\alpha}} (X_i - \omega) \right)\right]~,
\end{equation}
using which we can evaluate the variation of the measure $\Delta_S$
\begin{equation}
 \shift \Delta_S \left( \underline{X} \right) = \Delta_S \left( \underline{X} \right) \prod_{j \neq i}
  \frac{\sin \left( \tfrac{\pi}{\omega_{\alpha}} (X_i - X_j - \omega)
  \right)}{\sin \left( \tfrac{\pi}{\omega_{\alpha}} \left( X_i - X_j - \omega
  + M_{\mathrm{a}} \right) \right)} \cdot \frac{\sin \left(
  \tfrac{\pi}{\omega_{\alpha}} \left( X_j - X_i + M_{\mathrm{a}} \right)
  \right)}{\sin \left( \tfrac{\pi}{\omega_{\alpha}} (X_j - X_i) \right)}~.
\end{equation}
The variation of the antichirals then becomes
\begin{equation}
\begin{split}
\shift & \prod_{j=1}^{N}
  S_{2}\left(-X_{j}+\mu\right)^{-1}S_{2}\left(-X_{j}+\nu\right)^{-1}
  = \\
  & = \left[ \prod_{j=1}^{N}S_{2}\left(-X_{j}+\mu\right)^{-1}S_{2}\left(-X_{j}+\nu\right)^{-1}\right] 4\sin\left({\tfrac{{\pi}}{{\omega}_{{\alpha}}}}\left(X_{i}-\mu\right)\right)
     \sin\left({\tfrac{{\pi}}{{\omega}_{{\alpha}}}}\left(X_{i}-\nu\right)\right)\\
  &  =\left[ \prod_{j=1}^{N}S_{2}\left(-X_{j}+\mu\right)^{-1}S_{2}\left(-X_{j}+\nu\right)^{-1}\right] \left(-{\lambda}_{i,{\alpha}}\right)^{-1}\left( u_\alpha v_\alpha\right)^{\frac{1}{2}} P\left({\lambda}_{i,{\alpha}}\right)~,
\end{split}
\end{equation}
where for convenience we introduced $P(\lambda_{i,\alpha})$ as the quadratic polynomial defined by
\begin{equation}\label{eq:polynomial}
 P(\lambda) = 1+A\lambda+B\lambda^2~,
\end{equation}
with coefficients
\begin{equation}\label{eq:polynomial_coeffs}
\begin{split}
A &= -\left(u^{-1}_\alpha+v^{-1}_\alpha\right)
 = -\left( \mathe^{-\frac{2\pi\mathi\mu}{\omega_\alpha}} + \mathe^{-\frac{2\pi\mathi\nu}{\omega_\alpha}} \right) \\
B &= \left(u_\alpha v_\alpha\right)^{-1}
 = \mathe^{-\frac{2\pi\mathi(\mu+\nu)}{\omega_\alpha}}~.
 \end{split}
\end{equation}
Observe that this polynomial is of degree 2 precisely because we consider the inclusion of two antichiral fields. This fact will be important in Section~\ref{sec:recursive_solution} where we will find that the recursion relates correlators of degree $d$ with those of degree $d-1$ and $d-2$.

The variation of the CS and FI terms in \eqref{eq:J_definition} is
\begin{equation}
\begin{split}
 \shift \exp & \left(\sum_{j=1}^{N}\left[-{\frac{{\pi}{\mathi}{\kappa}_{2}}{{\omega}_{1}{\omega}_{2}}}X_{j}^{2} + {\frac{2{\pi}{\mathi}{\kappa}_{1}}{{\omega}_{1}{\omega}_{2}}}X_{j}\right]\right) = \\
  & \quad\quad =\exp\left(\sum_{j=1}^{N}\left[-{\frac{{\pi}{\mathi}{\kappa}_{2}}{{\omega}_{1}{\omega}_{2}}}X_{j}^{2}+{\frac{2{\pi}{\mathi}{\kappa}_{1}}{{\omega}_{1}{\omega}_{2}}}X_{j}\right]\right){\mathe}^{-{\frac{2{\pi}{\mathi}{\kappa}_{1}}{{\omega}_{{\alpha}}}}}q_{{\alpha}}^{-{\frac{{\kappa}_{2}}{2}}}\left(-{\lambda}_{i,{\alpha}}\right)^{{\kappa}_{2}}
\end{split}
\end{equation}
and finally that of the exponential of the times is
\begin{equation} \label{eq:second_exp_variation}
  \shift \exp \left( \sum_{s = 1}^{\infty} \tau_{s, \alpha} \sum_{j
  = 1}^N \lambda_{j, \alpha}^s \right) = \exp \left( \sum_{s = 1}^{\infty} \tau_{s, \alpha} \sum_{j
  = 1}^N \lambda_{j, \alpha}^s \right)\exp \left( \sum_{s = 1}^{\infty}
  \tau_{s, \alpha} (q_{\alpha}^{- s} - 1) \lambda_{i, \alpha}^s \right)~.
\end{equation}
We now evaluate the (RHS) in \eqref{eq:starting_eq} by inserting the variations computed in \eqref{eq:G_variation}-\eqref{eq:second_exp_variation} above, giving
\begin{equation}\label{eq:rhs}
\begin{split}
  \left(\mathrm{RHS}\right)
  = & \lbracket \sum_{i=1}^{N}{\frac{{\mathe}^{-{\frac{2{\pi}{\mathi}{\kappa}_{1}}{{\omega}_{{\alpha}}}}}q_{{\alpha}}^{-{\tfrac{{\kappa}_{2}}{2}}}\left(-{\lambda}_{i,{\alpha}}\right)^{{\kappa}_{2}-1}\left( u_\alpha v_\alpha\right)^{\frac{1}{2}}P\left({\lambda}_{i,{\alpha}}\right)}{z
  q^{-1}_{{\alpha}}{\lambda}_{i,{\alpha}}\left(1-z
  q^{-1}_{{\alpha}}
  {\lambda}_{i,{\alpha}}\right)}} \, \mathe^{\sum_{s=1}^{{\infty}}{\lambda}_{i,{\alpha}}^{s}\left(q_{{\alpha}}^{-s}-1\right){\tau}_{s,{\alpha}}}
  \left[\prod_{j{\neq}i}{\frac{{\lambda}_{i,{\alpha}}-t_{{\alpha}}{\lambda}_{j,{\alpha}}}{{\lambda}_{i,{\alpha}}-{\lambda}_{j,{\alpha}}}}\right]\rbracket_\tau\\
  = & -{\frac{{\mathe}^{-{\frac{2{\pi}{\mathi}{\kappa}_{1}}{{\omega}_{{\alpha}}}}}}{1-t_{{\alpha}}}} \Bigg\langle \sum_{i=1}^N {\frac{1}{2{\pi}{\mathi}}}\oint_{w={\lambda}_{i,{\alpha}}^{-1}}{\frac{{\mathd}w}{w}}\left[\sum_{n=-1}^{{\infty}}\left({\frac{z}{q_{{\alpha}}w}}\right)^{n}\right] \times \\
   & \times q^{-{\frac{{\kappa}_{2}}{2}}}_{{\alpha}} \left( -w\right)^{1-\kappa_2}\left( u_\alpha v_\alpha\right)^{\frac{1}{2}}P \left(\frac{1}{w}\right)
      \exp\left(\sum_{s=1}^{{\infty}}\frac{\left(q^{-s}_{{\alpha}}-1\right)}{w^s}{\tau}_{s,{\alpha}}\right) \prod_{j=1}^{N}{\frac{1-t_{{\alpha}}w{\lambda}_{j,{\alpha}}}{1-w{\lambda}_{j,{\alpha}}}}\Bigg\rangle_{\tau}~,
\end{split}
\end{equation}
where (following \cite{Lodin:2018lbz}) we have rewritten the expression inside of the average as a sum of residues at the points $w=\lambda_{i,\alpha}^{-1}$, for $w$ an auxiliary complex variable.
Now we use the fact that $w$ is a point on the Riemann sphere to move the contour in such a way that it encircles the poles at $w=0$ and $w=\infty$, \footnote{These are the only other poles of the integrand in \eqref{eq:rhs}.}
(with opposite orientation) instead of the poles at $w=\lambda_{i,\alpha}^{-1}$.

The (RHS) can thus be rewritten as
\begin{equation}\label{eq:F_definition}
\begin{split}
  \left(\mathrm{RHS}\right)
  = & {\frac{{\mathe}^{-{\frac{2{\pi}{\mathi}{\kappa}_{1}}{{\omega}_{{\alpha}}}}}}{1-t_{{\alpha}}}}
  \frac{1}{2\pi\mathi}\oint_{w=\{0,\infty\}} {\frac{\mathd w}{w}}\left[\sum_{n=-1}^{{\infty}}\left({\frac{z}{q_{{\alpha}}w}}\right)^{n}\right] \times \\
   & \times \underbrace{ q^{-{\frac{{\kappa}_{2}}{2}}}_{{\alpha}} \left( -w\right)^{1-\kappa_2}\left( u_\alpha v_\alpha\right)^{\frac{1}{2}}P \left(\frac{1}{w}\right)
      \exp\left(\sum_{s=1}^{{\infty}}\frac{\left(q^{-s}_{{\alpha}}-1\right)}{w^s}{\tau}_{s,{\alpha}}\right) \lbracket \prod_{j=1}^{N}{\frac{1-t_{{\alpha}}w{\lambda}_{j,{\alpha}}}{1-w{\lambda}_{j,{\alpha}}}}\rbracket_{\tau} }_{F(w)}~,
\end{split}
\end{equation}
where we are now able to bring the matrix model average inside of the $w$ integral.

First we compute the residue at $w=\infty$ in \eqref{eq:F_definition} by substituting $F(w)$ with its power series expansion
\begin{equation}
 F(w) = \sum_{n\in\mathbb{Z}} F_n w^n~,
\end{equation}
so that we obtain
\begin{equation}
\begin{split}\label{eq:residueinfty}
  \res{w=\infty}\left({\ldots}\right) = & {\frac{1}{2{\pi}{\mathi}}}\oint_{w=\infty}{\frac{{\mathd}w}{w}}\left[\sum_{n=-1}^{{\infty}}\left({\tfrac{z}{q_{{\alpha}}w}}\right)^{n}\right] F(w) \\
  = & - \sum_{n=-1}^\infty F_n \left(\frac{z}{q_\alpha}\right)^n~.
\end{split}
\end{equation}
In order to determine the coefficients $F_n$ we need to specify the value of the CS level $\kappa_2$. For the result to be non-vanishing we first of all require $\kappa_2 \leq 2$. Secondly, the choice of $\kappa_2$ has to be such that the residue at $w=\infty$ yields a term which can cancel the expectation value of $\lambda_{i,\alpha}^{-1}$ appearing in \eqref{eq:LHS}. The only consistent such choice (that does not introduce any higher negative powers of $\lambda_{i,\alpha}$) is 
\begin{equation}
\kappa_2 = 1,
\end{equation}
which will be used in what follows.
Being in a neighborhood of $w=\infty$ we can use the identity
\begin{equation}
\begin{split}
 \prod_{i=1}^{N}{\frac{1-t_{{\alpha}}w{\lambda}_{i,{\alpha}}}{1-w{\lambda}_{i,{\alpha}}}}
 = t_\alpha^N \exp\left(\sum_{s=1}^{{\infty}}w^{-s}{\frac{\left(1-t^{-s}_{{\alpha}}\right)}{s}}\sum_{i=1}^{N}{\lambda}_{i,{\alpha}}^{-s}\right)
\end{split}
\end{equation}
to rewrite the last term in $F(w)$ and we immediately see that all coefficients $F_n$ vanish for $n>0$ so that the only contributions to \eqref{eq:residueinfty} are those for $n=-1,0$. An explicit computation gives
\begin{align}
  \res{w=\infty}\left({\ldots}\right) & = -t_{{\alpha}}^{N}q^{-{\frac{1}{2}}}_{{\alpha}}\left( u_\alpha v_\alpha\right)^{\frac{1}{2}} \times \\
  & \times \left[{\frac{q_{{\alpha}}}{z}}\left(A\,\Z\left(\tautau\right)+\left(q^{-1}_{{\alpha}}-1\right){\tau}_{1,{\alpha}}\Z\left(\tautau\right)+\left(1-t^{-1}_{{\alpha}}\right)\sum_{i=1}^{N}\lbracket {\frac{1}{{\lambda}_{i,{\alpha}}}}\rbracket_\tau \right) + \Z\left(\tautau\right)\right]~. \notag
\end{align}
We now consider the other residue in \eqref{eq:F_definition}, namely the residue at $w=0$. We first rewrite the integral as
\begin{equation}
\begin{split}
  \res{w=0}\left({\ldots}\right) & = {\frac{1}{2{\pi}{\mathi}}}\oint_{w=0}{\frac{{\mathd}w}{w}}\left[\sum_{n=-1}^{{\infty}}\left({\tfrac{z}{q_{{\alpha}}w}}\right)^{n}\right]F\left(w\right) \\
  & = \sum_{n=-1}^{{\infty}}F_{n}\left({\tfrac{z}{q_{{\alpha}}}}\right)^{n} \\
  & = F\left(\tfrac{z}{q_\alpha}\right) - \underbrace{\sum_{n=2}^\infty F_{-n} \left(\tfrac{z}{q_\alpha}\right)^{-n}}_{\texttt{remainder}}~,
\end{split}
\end{equation}
where $F_n$ are now the coefficients of the power series expansion of $F(w)$ around $w=0$ and \texttt{remainder} is the part of the series which only contains negative powers of $z$ of degree less than $-1$.
Given that we are only interested in the $q$-Virasoro constraints for $n\geq -1$, such spurious terms can be neglected in the derivation of the main equation and therefore we will not be interested in writing their particular expression.

Next, as we are working in a small enough neighborhood of $w=0$, we can use the identity
\begin{equation}
\begin{split}
 \prod_{i=1}^{N}{\frac{1-t_{{\alpha}}w{\lambda}_{i,{\alpha}}}{1-w{\lambda}_{i,{\alpha}}}}
 = \exp\left(\sum_{s=1}^{{\infty}}w^{s}{\frac{\left(1-t^{s}_{{\alpha}}\right)}{s}}\sum_{i=1}^{N}{\lambda}_{i,{\alpha}}^{s}\right)
\end{split}
\end{equation}
to rewrite the residue as
\begin{align}
 \res{w=0}\left({\ldots}\right) & = q^{-{\tfrac{1}{2}}}_{{\alpha}}\left( u_\alpha v_\alpha\right)^{\tfrac{1}{2}} \times \\
 & \times P\left({\tfrac{q_{{\alpha}}}{z}}\right)\exp\left(\sum_{s=1}^{{\infty}}\frac{\left(1-q^{s}_{{\alpha}}\right)}{z^s}{\tau}_{s,{\alpha}}\right)\exp\left(\sum_{s=1}^{{\infty}}z^{s}{\frac{\left(1-t_{{\alpha}}^{s}\right)}{s q_{{\alpha}}^{s}}}{\frac{{\partial}}{{\partial}{\tau}_{s,{\alpha}}}}\right)\Z\left(\tautau\right) - \texttt{remainder} \notag
\end{align}
again using $\kappa_2=1$.

Finally, we can combine both residues and plug everything back into the original equation \eqref{eq:starting_eq}, to obtain
\begin{equation}
\begin{split}
  &
  t_{{\alpha}}^{N}\exp\left(\sum_{s=1}^{{\infty}}z^{s}{\frac{\left(1-t_{{\alpha}}^{-s}\right)}{s}} \, {\frac{{\partial}}{{\partial}{\tau}_{s,{\alpha}}}}\right)\Z\left(\tautau\right)+\\
  &
  +{\mathe}^{-{\frac{2{\pi}{\mathi}{\kappa}_{1}}{{\omega}_{{\alpha}}}}}q_{{\alpha}}^{-{\tfrac{1}{2}}}\left( u_\alpha v_\alpha\right)^{\frac{1}{2}}P\left({\tfrac{q_{{\alpha}}}{z}}\right)\exp\left(\sum_{s=1}^{{\infty}}z^{-s}\left(1-q^{s}_{{\alpha}}\right){\tau}_{s,{\alpha}}\right)\exp\left(\sum_{s=1}^{{\infty}}z^{s}{\frac{\left(1-t_{{\alpha}}^{s}\right)}{s
  q_{{\alpha}}^{s}}}{\frac{{\partial}}{{\partial}{\tau}_{s,{\alpha}}}}\right)\Z\left(\tautau\right)=\\
  = &
  \left(1+{\mathe}^{-{\frac{2{\pi}{\mathi}{\kappa}_{1}}{{\omega}_{{\alpha}}}}}t_{{\alpha}}^{N}q_{{\alpha}}^{-{\frac{1}{2}}}\left( u_\alpha v_\alpha\right)^{\frac{1}{2}}\right)\Z\left(\tautau\right)+{\frac{{\mathe}^{-{\frac{2{\pi}{\mathi}{\kappa}_{1}}{{\omega}_{{\alpha}}}}}t_{\alpha}^N q^{{\frac{1}{2}}}_{{\alpha}}\left( u_\alpha v_\alpha\right)^{\frac{1}{2}}}{z}}\left[\left(q^{-1}_{{\alpha}}-1\right){\tau}_{1,{\alpha}}+A\right]\Z\left(\tautau\right)+\\
  &
  +{\frac{1}{z}}(1-t_\alpha)\sum_{i=1}^{N}\lbracket{\frac{1}{{\lambda}_{i,{\alpha}}}}\rbracket_\tau\left(1-{\mathe}^{-{\frac{2{\pi}{\mathi}{\kappa}_{1}}{{\omega}_{{\alpha}}}}}q_{{\alpha}}t_\alpha^{N-1}q^{-{\frac{1}{2}}}_{{\alpha}}\left( u_\alpha v_\alpha\right)^{\frac{1}{2}}\right) +\texttt{remainder}~.
\end{split}
\end{equation}
In order to have the cancellation of the term $\lbracket1/\lambda_{i,\alpha}\rbracket_\tau$ as discussed above, we set the value of $\kappa_1$ accordingly\footnote{This can be compared to the value of $\kappa_1$ in \cite{Nedelin:2016gwu} where the additional terms $-\tfrac{\omega}{2}+\tfrac{\mu+\nu}{2}$ are generated by the inclusion of the fundamental chiral multiplets and we set {\it their} parameter $\alpha=0$ (not to be confused with our index $\alpha$).}
\begin{equation}
 \kappa_1 =
  \omega +M_{\mathrm{a}}(N-1)-\tfrac{\omega}{2}+\tfrac{\mu+\nu}{2}~,
\end{equation}
which leads to the constraint equation
\begin{equation} \label{eq:combined_constraint}
\boxed{
\begin{aligned}
  &
  t_{{\alpha}}^{N}\exp\left(\sum_{s=1}^{{\infty}}z^{s}{\frac{\left(1-t_{{\alpha}}^{-s}\right)}{s}} \, {\frac{{\partial}}{{\partial}{\tau}_{s,{\alpha}}}}\right)\Z\left(\tautau\right)+\\
  &
  +q_{{\alpha}}^{-1}t_{{\alpha}}^{1-N}P\left({\tfrac{q_{{\alpha}}}{z}}\right)\exp\left(\sum_{s=1}^{{\infty}}z^{-s}\left(1-q^{s}_{{\alpha}}\right){\tau}_{s,{\alpha}}\right)\exp\left(\sum_{s=1}^{{\infty}}z^{s}{\frac{\left(1-t_{{\alpha}}^{s}\right)}{s
  q_{{\alpha}}^{s}}}{\frac{{\partial}}{{\partial}{\tau}_{s,{\alpha}}}}\right)\Z\left(\tautau\right)=\\
  = &
  \left(1+q_{{\alpha}}^{-1}t_{{\alpha}}\right)\Z\left(\tautau\right)+{\frac{t_{{\alpha}}}{z}}\left[\left(q^{-1}_{{\alpha}}-1\right){\tau}_{1,{\alpha}}+A\right]\Z\left(\tautau\right)+\texttt{remainder}~.
\end{aligned}
}
\end{equation}
Thus, what we did is to rewrite the finite difference equation \eqref{eq:starting_eq} as a differential equation in the time variables $\tau_{s,\alpha}$ for the generating function $\Z(\tautau)$. Upon expanding this equation in powers of $z$ (for $n\geq -1$) we obtain a set of differential constraints which we can interpret as an explicit representation for the $q$-Virasoro algebra (see Section~\ref{sec:free_field}). From now on we will refer to \eqref{eq:combined_constraint} as the (combined) $q$-Virasoro constraints.

In Section~\ref{sec:recursive_solution} we will provide a recursive solution for this set of constraints.

\subsection{Free field representation}\label{sec:free_field}
Before attempting to solve equation \eqref{eq:combined_constraint} we want to provide an algebraic description of the constraints
and their relation to the representation theory of the $q$-Virasoro algebra as previously studied in \cite{Shiraishi:1995rp,Nedelin:2016gwu}. The generating function defined in Section~\ref{sec:gauge-theory-def} can be interpreted as a highest weight vector in a module over two commuting copies of the $q$-Virasoro algebra, one for each value of the label $\alpha$. Each copy of the algebra is generated by the operators $\hat{T}_{n,\alpha}$ for $n\in\mathbb{Z}$ with which we can express the $q$-Virasoro constraints as
\begin{equation}
 \hat{T}_{n,\alpha} \, \Z(\tautau) = 0, \quad n \geq 1
\end{equation}
expressing the condition that $\Z(\tautau)$ is indeed a highest weight vector annihilated by all the positive generators (the generators $\hat{T}_{0,\alpha}$ are diagonal on the highest weight and they give simple eigenvalue equations when acting on the generating function).

Notice that here we use the ``hatted'' notation $\hat{T}_{\alpha}(z)$ instead of the standard non-hatted current of \cite{Shiraishi:1995rp} to stress that the representation of the algebra is deformed by the introduction of the antichiral fundamental multiplets \eqref{eq:antichirals}, which amounts to the time shifts \eqref{eq:shifts}.

It is then customary to package the full set of generators into a stress tensor current $\hat{T}_{\alpha}(z)$ as
\begin{equation}
 \hat{T}_{\alpha}(z) := \sum_{n\in\mathbb{Z}} \hat{T}_{n,\alpha} z^n
\end{equation}
so that the constraints can be collectively rewritten as
\begin{equation} \label{eq:qVir_constraints}
 \hat{T}_{\alpha}(z) \Z(\tautau) = \mathrm{Pol}_\alpha(z)~,
\end{equation}
where $\mathrm{Pol}_\alpha(z)$ is a function whose power series expansion only contains non-positive powers of $z$. By expanding in powers of $z$ on both sides of the equation, one recovers the action of each of the generators of the algebra.

What we want to do now is to interpret equation \eqref{eq:combined_constraint} as a concrete representation for the algebraic identity \eqref{eq:qVir_constraints}.
In order to do that, we first introduce the following representation for the Heisenberg oscillators of \cite[Section~4]{Shiraishi:1995rp}
\begin{equation}\label{eq:heisengerg_oscillators}
\begin{aligned}
  \mathsf{a}_{s,\alpha} = (p^{s / 2}_{\alpha} q_{\alpha}^{- s}) \frac{\partial}{\partial \tau_{s,
  \alpha}}, \quad & \mathsf{a}_{- s,\alpha} = s \frac{1 - q_{\alpha}^s}{1 - t_{\alpha}^s} (p^{-
  s / 2}_{\alpha} q_{\alpha}^s) \tau_{s, \alpha}, \quad s \geq 1 \\
  & \mathsf{a}_{0,\alpha} = N
\end{aligned}
\end{equation}
satisfying the algebra
\begin{equation}
  [\mathsf{a}_{n,\alpha}, \mathsf{a}_{m,\alpha'}] = n \frac{1 - q_{\alpha}^{| n |}}{1 - t_{\alpha}^{| n |}}
  \delta_{n + m, 0}\delta_{\alpha,\alpha'}~.
\end{equation}
By using this explicit free field representation and also introducing the function $\psi_\alpha(z)$ as
\begin{equation}
  \psi_{\alpha} (z) \assign p_{\alpha}^{- 1 / 2} \exp \left( \sum_{s =
  1}^{\infty} z^{- s} \frac{(1 - q_{\alpha}^s)}{(1 + p^s_{\alpha})} \tau_{s,
  \alpha} \right)
\end{equation}
 we are able to rewrite \eqref{eq:combined_constraint} as
\begin{equation} \label{eq:free_field}
  \psi_{\alpha} (z) \hat{T}_{\alpha} (z) \Z \left( \tautau \right) =
  (1 + p_{\alpha}^{- 1}) \Z \left( \tautau \right) +
  \frac{t_{\alpha}}{z} \left[(q^{- 1}_{\alpha} - 1) \tau_{1, \alpha} +
  A\right] \Z \left( \tautau
  \right) + \texttt{remainder}~,
\end{equation}
where the current $\hat{T}_{\alpha}(z)$ takes the form of the differential operator
\begin{equation}\label{eq:current}
\begin{split}
  \hat{T}_{{\alpha}}\left(z\right)= &
  p_{{\alpha}}^{1/2}\exp\left(-\sum_{s=1}^{{\infty}}z^{-s}{\frac{\left(1-q_{{\alpha}}^{s}\right)}{\left(1+p^{s}_{{\alpha}}\right)}}
  {\tau}_{s,{\alpha}}\right)\exp\left(-\sum_{s=1}^{{\infty}}z^{s}{\frac{\left(1-t_{{\alpha}}^{s}\right)}{s t^s_\alpha}} \, {\frac{{\partial}}{{\partial}{\tau}_{s,{\alpha}}}}\right)t^{N}_{{\alpha}}+\\
  &
  +P\left({\tfrac{q_{{\alpha}}}{z}}\right)p_{{\alpha}}^{-1/2}\exp\left(\sum_{s=1}^{{\infty}}z^{-s}{\frac{\left(1-q^{s}_{{\alpha}}\right)}{\left(1+p^{s}_{{\alpha}}\right)}}p^{s}_{{\alpha}}{\tau}_{s,{\alpha}}\right)\exp\left(\sum_{s=1}^{{\infty}}z^{s}{\frac{\left(1-t_{{\alpha}}^{s}\right)}{s
  q_{{\alpha}}^{s}}}{\frac{{\partial}}{{\partial}{\tau}_{s,{\alpha}}}}\right)t^{-N}_{{\alpha}}~,
\end{split}
\end{equation}
and \texttt{remainder} is identified with the part of $\psi_\alpha(z)\mathrm{Pol}_\alpha(z)$ with powers of $z$ of degree less than $-1$.

Comparing \eqref{eq:current} with the formula for the current of \cite{Shiraishi:1995rp} we observe that the only difference is the multiplicative factor $P\left(q_\alpha/z\right)$ appearing in front of the second term. This deformation is due to the presence of the two anti-fundamental flavors of masses $\mu$ and $\nu$ on which the polynomial $P$ depends through the coefficients $A,B$. As a direct consequence we observe that the constraint equation for $\hat{T}_{-1,\alpha}$ is also modified as
\begin{equation}
\begin{split}
  \hat{T}_{-1,{\alpha}}\Z\left(\tautau\right) &
  =p_{{\alpha}}^{{\frac{1}{2}}}\left[t_{\alpha}\left[\left(q^{-1}_{{\alpha}}-1\right){\tau}_{1,{\alpha}}+A\right]-{\frac{\left(1-q_{{\alpha}}\right)}{\left(1+p_{{\alpha}}\right)}}
  {\tau}_{1,{\alpha}}\left(1+p_{{\alpha}}^{-1}\right)\right]\Z\left(\tautau\right)\\
  &
  =A\, q_{\alpha} \, p_{{\alpha}}^{-1/2}\Z\left(\tautau\right)~.
\end{split}
\end{equation}
The equation for $\hat{T}_{0,\alpha}$ instead does not depend on the deformation and gives the usual eigenvalue equation
\begin{equation}
  \hat{T}_{0, \alpha} \Z \left( \tautau \right) = \left(
  p_{\alpha}^{\frac{1}{2}} + p_{\alpha}^{- \frac{1}{2}} \right) \Z \left(
  \tautau \right)
\end{equation}
that we expect from a highest weight module representation.
%
%
%
\section{Recursive solution}
\label{sec:recursive_solution}
%
%
%
The goal of this section is to solve the $q$-Virasoro constraints in \eqref{eq:combined_constraint}, where by solve we mean to recursively determine all the normalized correlators $\C_{\ell_1 \dots  \ell_{n} ; k_1 \dots  k_{m}}$ of this model. The correlators are defined as 
\begin{equation}
\begin{split}
 \C_{\ell_1 \dots  \ell_{n} ; k_1 \dots  k_{m}} \assign &
 \lbracket p_{\ell_1}(\left\{ \lambda_{i,1} \right\}) \dots p_{\ell_n}(\left\{ \lambda_{i,1} \right\}) \, p_{k_1}(\left\{ \lambda_{i,2} \right\}) \dots p_{k_m}(\left\{ \lambda_{i,2} \right\}) \rbracket_{\tau=0} \\
 =& \left[ \frac{\partial^n}{\partial \tau_{\ell_1,1}\dots \partial \tau_{\ell_n,1}}  \frac{\partial^m}{\partial \tau_{k_1,2}\dots \partial \tau_{k_m,2}}   \Z\left(\tautau\right) \right]_{\underline{\tau}_1,\underline{\tau}_2=0}~.
\end{split}
\end{equation}

We assume that the partition function admits the formal power series expansion:
\begin{align}\label{eq:partition_fn_expansion}
\Z\left(\tautau\right) & =  \sum_{d_1 =0}^\infty \, \sum_{d_2 =0}^\infty \, \Z_{d_1,d_2}\left(\tautau\right) \notag\\
 & =  \sum_{d_1,d_2 =0}^\infty \, \sum_{n,m =0}^\infty \frac{1}{n!} \frac{1}{m!} \, \sum_{\ell_1 + \dots + \ell_{n} = d_1} \, \sum_{k_1 + \dots + k_{m} = d_2} \C_{\ell_1 \dots \ell_{n} ; k_1 \dots k_{m}} \, \tau_{\ell_1,1} \dots \tau_{\ell_{n},1} \, \tau_{k_1,2} \dots \tau_{k_{m},2}~,
\end{align}
where $\Z_{d_1,d_2}\left(\tautau\right)$ has degree $d_1$ and $d_2$ with respect to the operators $\sum_{d_1=1}^\infty d_1 \tau_{d_1,1} \frac{\partial}{\partial \tau_{d_1,1}}$ and $\sum_{d_2=1}^\infty d_2 \tau_{d_2,2} \frac{\partial}{\partial \tau_{d_2,2}}$, respectively.
We then use the definition \eqref{eq:symm-schur-poly} of the symmetric Schur polynomial $s_{\{m\}}(p_1,\dots,p_m)$ for a given symmetric partition $\{m\}$ in order to extract the coefficient of $z^m$, $m=-1,0,1,\dots$ in \eqref{eq:combined_constraint}. When doing so, we only consider terms of a particular degree $d_{1}$ in times $\tau_{\ell,1}$ and degree $d_{2}$ in times $\tau_{k,2}$.

By inserting formula \eqref{eq:partition_fn_expansion} into the $q$-Virasoro constraint \eqref{eq:combined_constraint} and choosing $\alpha=1$ for definiteness, we get
\begin{align}\label{eq:recursion1}
  & t^N_1 \sum_{\ell=0}^{d_1} s_{\{\ell\}} \left(\left\{p_s=-s(1-q^s_1)\tau_{s,1}\right\}\right) s_{\{\ell+m\}} \left(\left\{p_s=(1-t^{-s}_1)\frac{\partial}{\partial\tau_{s,1}}\right\}\right) \Z_{d_1+m,d_{2}}\left(\tautau\right) \notag \\
+ & q^{-1}_1 t^{1-N}_1 s_{\{m\}}\left(\left\{p_s=\frac{1-t_1^s}{q^s_1}\frac{\partial}{\partial\tau_{s,1}}\right\}\right)\Z_{d_1+m,d_{2}}\left(\tautau\right) \notag \\
+ & A \, t^{1-N}_1 s_{\{m+1\}}\left(\left\{p_s=\frac{1-t_1^s}{q^s_1}\frac{\partial}{\partial\tau_{s,1}}\right\}\right)\Z_{d_1+m+1,d_{2}}\left(\tautau\right) \\
+ & B \, q_1 t^{1-N}_1 s_{\{m+2\}}\left(\left\{p_s=\frac{1-t_1^s}{q^s_1}\frac{\partial}{\partial\tau_{s,1}}\right\}\right)\Z_{d_1+m+2,d_{2}}\left(\tautau\right) \notag \\
= & \delta_{m,0} (1+q^{-1}_1 t_1)\Z_{d_1,d_{2}}\left(\tautau\right)
- \delta_{m,-1} \left( (1-q_1)\tau_{1,1}\Z_{d_1-1,d_{2}}\left(\tautau\right) - A\, t_1 \Z_{d_1,d_{2}}\left(\tautau\right) \right) \notag~.
\end{align}
The corresponding equation for $\alpha=2$ is completely analogous but it is $d_2$ that is shifted by $m$.

Given any two partitions $\rho=\{\rho_1,\dots,\rho_\bullet\}$ and $\sigma=\{\sigma_1,\dots,\sigma_\star\}$, where $\rho_\bullet$ and $\sigma_\star$ indicate the last components, we wish to compute the correlator $\C_{\rho;\sigma} \equiv \C_{\rho_1 \dots\rho_\bullet;\sigma_1\dots\sigma_\star}$,
with the identifications $m+2=\rho_\bullet$, $d_1+2=|\rho|$ and $d_2=|\sigma|$.
To extract the correlator we are interested in, we apply the operator
\begin{equation}
\frac{\partial^{\bullet - 1}}{\partial \tau_{\rho_1,1} \dots \partial \tau_{\rho_{\bullet-1},1}} \frac{\partial^{\star}}{\partial \tau_{\sigma_1,2} \dots \partial \tau_{\sigma_{\star},2}} \Bigg{|}_{\underline{\tau}_1, \underline{\tau}_2 = 0}
\end{equation}
to \eqref{eq:recursion1}, namely we differentiate with respect to the corresponding combination of times and then set all of them to zero.
Finally, we obtain
\begin{equation}\label{eq:recursion}
\begin{split}
  & - B \, q_1 t^{1-N}_1 \left( \frac{(1 - t_1^{\rho_\bullet})}{q_1^{\rho_\bullet}\rho_\bullet} \right)
    \C_{\rho_1 \dots \rho_\bullet;\sigma} \\
= & + B \, q_1 t^{1-N}_1 \sum_{\left\{\substack{\gamma \;\text{s.t.}\,|{\gamma}| = \rho_\bullet \\ l({\gamma}) \geq 2}\right\}}
    \frac{1}{|\mathrm{Aut}(\gamma)|} \left ( \prod_{a \in {\gamma}} \frac{(1 - t_1^a)}{q_1^a a} \right )
    \C_{\rho_1 \dots \rho_{\bullet - 1}\gamma_1 \dots \gamma_{\bullet} ;\sigma} \\
  & + A \, t^{1-N}_1 \sum_{\{\gamma\;\text{s.t.}\,|{\gamma}| = \rho_\bullet -1\}}
    \frac{1}{|\mathrm{Aut}({\gamma})|} \left ( \prod_{a \in {\gamma}} \frac{(1 - t_1^a)}{q_1^a a} \right )
    \C_{\rho_1 \dots \rho_{\bullet - 1} \gamma_1 \dots \gamma_{\bullet};\sigma} \\
  & + q^{-1}_1 t^{1-N}_1 \sum_{\{\gamma\;\text{s.t.}\,|{\gamma}| = \rho_\bullet-2\}}
    \frac{1}{|\mathrm{Aut}({\gamma})|} \left ( \prod_{a \in {\gamma}} \frac{(1 - t_1^a)}{q_1^a a} \right )
    \C_{\rho_1 \dots \rho_{\bullet - 1}\gamma_1 \dots \gamma_{\bullet} ;\sigma} \\
  & +t^N_1 \sum_{\eta \subseteq \rho \setminus \rho_\bullet}
    \left ( \prod_{a \in \eta} (q_1^a-1) \right ) \sum_{\{\gamma \;\text{s.t.}\,|{\gamma}| = |\eta| + \rho_\bullet - 2\}}
    \frac{1}{|\mathrm{Aut}({\gamma})|} \left ( \prod_{a \in {\gamma}} \frac{(1 - t_1^{-a})}{a} \right )
    \C_{\rho \setminus \{\rho_\bullet, \eta\} \gamma_1 \dots \gamma_{\bullet};\sigma}
     \\
  & -\delta_{\rho_\bullet, 2}  (1+q^{-1}_1 t_1)\C_{\rho_1 \dots \rho_{\bullet - 1};\sigma}
+\delta_{\rho_\bullet, 1} \left( (1-q_1)((\#_\rho 1) - 1)\C_{\rho_1 \dots \rho_{\bullet - 2};\sigma} - A\, t_1 \C_{\rho_1 \dots \rho_{\bullet - 1};\sigma}\right)~,
\end{split}
\end{equation}
which can be used to determine the correlator $\C_{\rho;\sigma}$ in terms of the lower order correlators on the right hand side.
Here $\gamma$ and $\eta$ are partitions and we used the formulas of Appendix~\ref{sec:schur} to rewrite explicitly the symmetric Schur polynomials $s_{\{m\}}(\{p_k\})$ in \eqref{eq:partition_fn_expansion}.
The sum over subsets $\eta\subseteq\rho \setminus \rho_\bullet$ means that we are summing over all the sub-partitions $\eta$ of $\rho$ not containing its last part $\rho_\bullet$.
An analogous equation holds for simplifying the multi-index $\sigma$ of the correlator.

Every correlator is uniquely determined by repeated application of \eqref{eq:recursion}. Therefore, if a non-trivial solution to the over-determined system of $q$-Virasoro constraints \eqref{eq:combined_constraint} exists, then it must be given by \eqref{eq:recursion}. While the proof of consistency of \eqref{eq:combined_constraint} is out of the scope of this paper, we did check (up to degree 6) that it is indeed consistent.

The recursion in \eqref{eq:recursion} treats the two copies $\alpha=1,2$ separately. 
From this observation, one might be lead to believe that the correlators factorize, although this is not obvious from the form of the partition function \eqref{eq:partitionfunction}. We now prove that this is indeed the case. First of all suppose that all correlators for partitions up to a certain order $d$ factorize into the product of correlators for $\alpha=1$ and $\alpha=2$ as
\begin{equation} \label{eq:factorization}
\C_{\rho;\sigma} = \C_{\rho;\emptyset}\C_{\emptyset;\sigma} = \C_\rho^{1} \C_\sigma ^{2}~.
\end{equation}
Then, if we consider a partition $\rho$ of order $d+1$, all correlators in the right hand side of the recursion formula \eqref{eq:recursion} factorize as per our assumption and therefore we can collect a common factor of $\C_{\emptyset;\sigma}$. This immediately implies that the ratio $\C_{\rho;\sigma}/\C_{\emptyset;\sigma}$ can be written entirely in terms of correlators of the form $\C_{\rho';\emptyset}$ for $\rho'$ some partition of order at most $d$. In other words, the correlator on the left hand side (which is of order $d+1$) must also factorize as prescribed in \eqref{eq:factorization}. Finally we need to show that the initial data of the recursion factorizes as well. However, the only correlator needed to fix this data is the trivial correlator $\C_{\emptyset;\emptyset}$ whose value can be set to 1 as a choice of overall normalization 
\begin{equation}
\C_{\emptyset, \emptyset} = 1~,
\end{equation}
 which corresponds to looking at the normalized correlators. 
The factorization of all correlators then follows by induction, as well as that of all Wilson loop expectation values
\begin{equation}
 \lbracket \mathrm{WL}_\rho^{(1)} \mathrm{WL}_\sigma^{(2)} \rbracket = \lbracket \mathrm{WL}_\rho^{(1)} \rbracket \lbracket \mathrm{WL}_\sigma^{(2)} \rbracket
\end{equation}
for any pair of partitions $\rho,\sigma$. While non-obvious from first principles, this factorization property is similar to the holomorphic blocks factorization of \cite{Pasquetti:2011fj,Beem:2012mb}.
\\
\\
With the help of the recursion we find the first few correlators
\begin{align}
\C_{1}^{\alpha} & = -\frac{A(t_\alpha^N-1)}{B(t_\alpha-1)} ~, \\
\C_{2}^{\alpha} & = \frac{\left(t_\alpha^N-1\right) \left(A^2 t_\alpha \left(t_\alpha^N+1\right)+B \left((q_\alpha-1) t_\alpha^{N+1}+(q_\alpha+1) t_\alpha^N-2 t_\alpha\right)\right)}{B^2 t_\alpha \left(t_\alpha^2-1\right)} \\
\C_{1,1}^{\alpha} & = \frac{\left(t_\alpha^N-1\right) \left(A^2 t_\alpha \left(t_\alpha^N-1\right)+B (q_\alpha-1) (t_\alpha-1) t_\alpha^N\right)}{B^2 (t_\alpha-1)^2 t_\alpha} ~, \\
\C_{3}^{\alpha} & = -\frac{A \left(t_\alpha^N-1\right) }{B^3 t_\alpha^2 \left(t_\alpha^3-1\right)} \Bigg[A^2 t_\alpha^2 \left(t_\alpha^{2 N}+t_\alpha^{N}+1\right)+B \Big(\left(q_\alpha^2+q_\alpha+1\right) t_\alpha^{2 N}+\left(q_\alpha^2+q_\alpha+1\right) t_\alpha^{2 N+1}+ \notag \\
& \quad \quad \quad +\left(q_\alpha^2+q_\alpha-2\right) t_\alpha^{2 N+2}-3 t_\alpha^{N+2}-3 t_\alpha^2\Big)\Bigg] ~,\\
\C_{2,1}^{\alpha} & = -\frac{A \left(t_\alpha^N-1\right) }{B^3 (t_\alpha-1)^2 t_\alpha^2 (t_\alpha+1)} \Bigg[A^2 t_\alpha^2 \left(t_\alpha^{2 N}-1\right)+B \Big(-\left(q_\alpha^2-1\right) t_\alpha^{2 N}+\left(q_\alpha^2+q_\alpha-2\right) t_\alpha^{2 N+2}+\notag \\
& \quad \quad \quad -(q_\alpha+1) t_\alpha^{N+1} -(q_\alpha+1) t_\alpha^{N+2}+(q_\alpha+1) t_\alpha^{2 N+1}+2 t_\alpha^2\Big)\Bigg] ~,
\end{align} 
\begin{align}
\C_{1,1,1}^{\alpha} & = -\frac{A \left(t_\alpha^{N}-1\right) \left(A^2 t_\alpha^2 \left(t_\alpha^{N}-1\right)^2+B (q_\alpha-1) (t_\alpha-1) t_\alpha^{N} \left((q_\alpha+2) t_\alpha^{N+1}-(q_\alpha-1) t_\alpha^{N}-3 t_\alpha\right)\right)}{B^3 (t_\alpha-1)^3 t_\alpha^2} ~,
\end{align}
where we recall that $A$ and $B$ are the coefficients of the polynomial $P(\lambda)$ given in \eqref{eq:polynomial_coeffs}.

For the first few correlators of Schur polynomials we manifestly get 
\begin{align}
 \lbracket s_{\{1\}}(\left\{ \lambda_{i,\alpha} \right\}) \rbracket & = \C_1^\alpha
 = -\frac{A(t_\alpha^N-1)}{B(t_\alpha-1)} ~, \\
\lbracket s_{\{2\}}(\left\{ \lambda_{i,\alpha} \right\}) \rbracket & = \frac{\C_{1,1}^\alpha+\C_2^\alpha}{2} = \\
 & = \frac{\left(t_\alpha^{N}-1\right) \left(A^2 t_\alpha \left(t_\alpha^{N+1}-1\right)+B (t_\alpha-1) \left((q_\alpha-1) t_\alpha^{N+1}+q_\alpha t_\alpha^{N}-t_\alpha\right)\right)}{B^2 (t_\alpha-1)^2 t_\alpha (t_\alpha+1)} ~,\notag\\
 \lbracket s_{\{1,1\}}(\left\{ \lambda_{i,\alpha} \right\}) \rbracket & = \frac{\C_{1,1}^\alpha-\C_2^\alpha}{2}= \frac{\left(t_\alpha-t_\alpha^{N}\right) \left(t_\alpha^{N}-1\right) \left(B (t_\alpha-1)-A^2 t_\alpha\right)}{B^2 (t_\alpha-1)^2 t_\alpha (t_\alpha+1)} ~, \\
\lbracket s_{\{3\}}(\left\{ \lambda_{i,\alpha} \right\}) \rbracket & = \frac{\C_{1,1,1}^\alpha+ 3\C_{2,1}^\alpha+2\C_3^\alpha}{6} = \\
& = -\frac{A \left(t_\alpha^{N}-1\right) }{B^3 (t_\alpha-1)^3 t_\alpha^2 (t_\alpha+1) \left(t_\alpha^2+t_\alpha+1\right)}
\Bigg[ A^2 \left(-t_\alpha^{N+3}-t_\alpha^{N+4}+t_\alpha^{2 N+5}+t_\alpha^2\right) \notag\\
& +B (t_\alpha-1) \Big(-q_\alpha^2 t_\alpha^{2 N}+\left(q_\alpha^2+2 q_\alpha-1\right) t_\alpha^{2 N+3}+ \notag \\
& +\left(q_\alpha^2+q_\alpha-2\right) t_\alpha^{2 N+4}-q_\alpha t_\alpha^{N+1}+(1-2 q_\alpha) t_\alpha^{N+2}-2 q_\alpha t_\alpha^{N+3}-(q_\alpha+1) t_\alpha^{N+4}+\notag \\
& -(q_\alpha-1) q_\alpha t_\alpha^{2 N+1}+2 q_\alpha t_\alpha^{2 N+2}+t_\alpha^3+2 t_\alpha^2\Big)\Bigg] ~, \notag \\
\lbracket s_{\{2,1\}}(\left\{ \lambda_{i,\alpha} \right\}) \rbracket & = \frac{\C_{1,1,1}^\alpha-\C_3^\alpha}{3} =\\
& = -\frac{A \left(t_\alpha^{N}-1\right) \left(t_\alpha^{N}-t_\alpha\right) }{B^3 (t_\alpha-1)^3 t_\alpha^2 \left(t_\alpha^2+t_\alpha+1\right)} \times \notag\\
& \times \Big [A^2 t_\alpha^2 \left(t_\alpha^{N+1}-1\right)+B (t_\alpha-1) \left((q_\alpha-1) t_\alpha^{N+1}+(q_\alpha-2) t_\alpha^{N+2}+q t_\alpha^{N}-t_\alpha^2+t_\alpha\right)\Big ] ~, \notag \\
\lbracket s_{\{1,1,1\}}(\left\{ \lambda_{i,\alpha} \right\}) \rbracket & = \frac{\C_{1,1,1}^\alpha- 3\C_{2,1}^\alpha+2\C_3^\alpha}{6} = \\
& = -\frac{A \left(t_\alpha^{N}-1\right) \left(t_\alpha^{2 N}-t_\alpha^{N+1}-t_\alpha^{N+2}+t_\alpha^3\right) \left(A^2 t_\alpha^2+B\left(-2 t_\alpha^2+t_\alpha+1\right)\right)}{B^3 (t_\alpha-1)^3 t_\alpha^2 (t_\alpha+1) \left(t_\alpha^2+t_\alpha+1\right)}~, \notag
\end{align}
which can be translated to expectation values of Wilson loop operators via the relation
\begin{equation}
\langle \mathrm{WL}_{\rho}^{(\alpha)}\rangle = \lbracket s_{\rho}(\left\{ \lambda_{i,\alpha} \right\}) \rbracket~.
\end{equation}
Computation of any other Wilson loop expectation value goes through in exactly the same way.

In the next section we study several limits of the recursive solution outlined above.
%
%
%
\section{Limits} \label{sec:limits}
%
%
%
There are several interesting limits that can be taken starting from the result obtained in the previous section. 

\subsection{Round sphere}
Firstly, one can consider the round sphere limit which corresponds to letting $\omega_1=\omega_2$, or equivalently $q_\alpha \rightarrow 1$ for both values of $\alpha$.
In this case the value of $t=t_1=t_2=\mathe^{2\pi\mathi M_\mathrm{a}}$ is kept fixed and the $q$-difference equation \eqref{eq:combined_constraint} takes the simpler form:
\begin{equation}\label{eq:constraint_round_sphere}
\begin{aligned}
  &
  t^{-\tfrac{1}{2}+N}\exp\left(\sum_{s=1}^{{\infty}}z^{s}{\frac{\left(1-t^{-s}\right)}{s}} {\frac{{\partial}}{{\partial}{\tilde{\tau}}_{s}}}\right)\Z_{S^3}\left(\underline{\tilde{\tau}}\right)
  + t^{\tfrac{1}{2}-N}P\left({\tfrac{1}{z}}\right)\exp\left(\sum_{s=1}^{{\infty}}z^{s}{\frac{\left(1-t^{s}\right)}{s
  }}{\frac{{\partial}}{{\partial}{\tilde{\tau}}_{s}}}\right)\Z_{S^3}\left(\underline{\tilde{\tau}}\right)=\\
  & \quad\quad\quad =
  \left(t^{-\tfrac{1}{2}}+t^{\tfrac{1}{2}}\right)\Z_{S^3}\left(\underline{\tilde{\tau}}\right)+A\,{\frac{t^{\tfrac{1}{2}}}{z}}\Z_{S^3}\left(\underline{\tilde{\tau}}\right)+\texttt{remainder}~,
\end{aligned}
\end{equation}
where, since there is no distinction between the two copies $\alpha=1$ and $\alpha=2$, we can define a single set of time variables $\tilde{\tau}_s=\tau_{s,1}+\tau_{s,2}$.
The partition function $\Z_{S^3}\left(\underline{\tilde{\tau}}\right)$ of the gauge theory on the round sphere is a solution of this equation.
Observe also that in this case the two copies of the $q$-Virasoro algebra collapse to the same giving rise to just one set of constraints.

Furthermore we note that in this limit the shift in the time variables, which can be used to reproduce the contribution of the fundamental antichiral multiplets, becomes singular (see \eqref{eq:shifts}). This is also evident from the fact that in the round sphere limit we have $\mathrm{Im}(\omega_1/\omega_2)=0$ which is the region where the double sine cannot be factorized into a product of $q$-shifted factorials (see \eqref{eq:S2factorization}) and therefore \eqref{eq:antichirals} does not hold. 

Besides, in this limit the current in \eqref{eq:current} is such that it only contains {\it derivatives} in times. This can be seen from the observation that only half of the Heisenberg oscillators in \eqref{eq:heisengerg_oscillators} will remain, namely the positive modes. This renders the algebra of both the Heisenberg oscillators and the $q$-Virasoro generators abelian. We remark that this also corresponds to the Frenkel-Reshetikhin limit $W_{1,t}$ of \cite{frenkel1996}.

\subsection{Gaussian matrix model}
Secondly, by making the specific choice of masses such that $v_\alpha = -u_\alpha = (1-q_\alpha)^{-1/2}$, the contributions of the fundamental antichirals in \eqref{eq:antichirals} simplify to $q$-Pochhammer symbols using \eqref{eq:S2factorization}. For instance, for $\alpha=1$
\begin{equation}\label{eq:qGaussianMasses}
v_1 = -u_1 \quad \Rightarrow \quad   \nu = \mu \pm \frac{\omega_1}{2}~,
\end{equation}
where the ambiguity in the sign originates from interpreting the minus sign as an exponent. It can be noted that this depends only on $\omega_1$ and not $\omega_2$, and so we expect that this particular choice can only be made for one value of $\alpha$ at a time.
By setting the masses as in \eqref{eq:qGaussianMasses} the contribution of the fundamentals introduces a $q$-exponent factor in the first set of variables $\lambda_{i,1}$ of the form
\begin{equation}\label{eq:qExp}
\left(\lambda_{i,1}^{2} q_1^2 (1-q_1);q_1^2\right)_\infty~,
\end{equation}
which reproduces the potential of the $(q,t)$-Gaussian\footnote{The ``Gaussian'' in the name comes from the limit $q_1 \rightarrow 1$, in which the $q$-Pochhammer symbol \eqref{eq:qExp} becomes the standard Gaussian exponent.} model studied in \cite{Lodin:2018lbz}.
The dependence on the second set of variables $\lambda_{i,2}$ however, cannot be put in the form of a $q$-exponent function for the same values of masses.

\subsection{$U(1)$ gauge theory}
Thirdly, one can consider the $N = 1$ limit of the localized partition function on $S_b^3$, which is the case originally studied in \cite{Pasquetti:2011fj}.  The theory becomes that of a single $U(1)$ gauge group with two (anti)fundamental flavors while the adjoint multiplet becomes a singlet and decouples. In this case the integral in the partition function becomes 1-dimensional and the measure $\Delta_S \left( \underline{X} \right)$ in \eqref{eq:partitionfunction} is simply $1$. As a consequence of this simplification, the correlators in this limit do not depend on the adjoint chiral mass $M_{\mathrm{a}}$ and therefore $t_\alpha$ does not appear in the formulas. As is expected for a rank 1 matrix model, all correlators associated to partitions of the same degree become equal to each other. For example we have
\begin{equation}
\begin{aligned}
\left. \C_{1}^{\alpha} \right|_{N = 1} &= -\frac{A}{B}~, \\
\left. \C_{2}^{\alpha} \right|_{N = 1} = \left. \C_{1,1}^{\alpha} \right|_{N = 1} &= \frac{A^2+B (q_\alpha-1)}{B^2}~, \\
\left. \C_{3}^{\alpha} \right|_{N = 1} = \left. \C_{2,1}^{\alpha} \right|_{N = 1} = \left. \C_{1,1,1}^{\alpha} \right|_{N = 1} &= -\frac{A \left(A^2+B \left(q_\alpha^2+q_\alpha-2\right)\right)}{B^3}~.
\end{aligned}
\end{equation}
Moreover, one finds that all Schur polynomials vanish identically except for those associated to completely symmetric representations, i.e. to those partitions of length 1, so that we have
\begin{equation}
 \left. \lbracket s_{\{m\}}(\left\{ \lambda_{i,\alpha} \right\}) \rbracket \right|_{N=1} = \left. \C_m^\alpha\right|_{N=1}
\end{equation}
for all $m\geq 0$ where $m$ is the $U(1)$ charge of the representation.
%
%
%
\section{Conclusion} \label{sec:conclusion}
%
%
%
 In this paper we developed an explicit algorithm that allows one to calculate
 supersymmetric Wilson loop averages in $\mathcal{N}=2$ YM-CS theory coupled to specific matter on a squashed sphere $S^3_b$ for any value of $b$, including $b=1$.
 The heart of the algorithm is the Ward identities \eqref{eq:combined_constraint} for the corresponding matrix model,
 which produce the recursive equation \eqref{eq:recursion}
 that expresses power sum monomial correlators through simpler ones.

 Due to  technical reasons we restrict our attention
 to specific values of CS level $\kappa_2$  and FI parameter $\kappa_1$.
 Our current understanding suggests that such technical limitations reflect an underlying deeper conceptual difficulty
 related to the possibility to solve the matrix model only around a specific ``expansion point'',
 similarly to how the usual Hermitian matrix model is completely solvable by expanding around a Gaussian potential.
 Extending the present result to generic values of CS level and FI parameter is certainly an interesting question which
 requires further investigation.

 The present analysis is the natural continuation of the works \cite{Nedelin:2016gwu} and \cite{Lodin:2018lbz}. 
 In principle the present work could be extended to 3d ${\cal N}= 2$ unitary quiver gauge theories (which include the ABJ(M) model
 \cite{Aharony:2008ug,Aharony:2008gk} and its deformations) 
 as described formally in \cite{Nedelin:2016gwu} 
 with the Ward identities related to  $W_{q,t}(\Gamma)$ algebras of \cite{Kimura:2015rgi}. 
 The main problem is how to disentangle the algebraic and analytical issues in this generalization. 
 Within the formal free field theory representation all quiver theories can be treated uniformly. However we do not know how 
 to analyse $q$-difference operators, poles and contours in a uniform way.
 At the moment we can only do it model by model and it is a quite time-consuming analysis.  

 In a larger context it would be nice to understand how to solve $q$-Virasoro constraints in a more general situation.
 The generating functions for 3d and 5d gauge theories are naturally related to $q$-Virasoro constraints
 (and its relevant deformations \cite{Mironov:2016yue}). For example, the Nekrasov generating function  can be thought of
 as ``$N = \infty$'' $(q,t)$-deformed matrix models, see the Kimura-Pestun representation of Nekrasov function \cite{Kimura:2015rgi}.

 Last but not least, while from the point of view of $U(N)$ gauge theory Schur polynomials are the objects to consider,
 from a purely matrix model point of view a  different kind of polynomials -- the Macdonald polynomials -- are much more natural.
 They are the orthogonal polynomials for the $(q,t)$-Vandermonde measure,
 and their average is actually a very simple factorized expression \cite{Cordova:2016jlu,Morozov:2018eiq}
 -- a property known as ``character expansion'' \cite{Morozov:2018eiq}.
 Therefore, understanding the gauge theory significance of Macdonald polynomials
 is an important problem for future research.

{\section*{Acknowledgments}
 We thank Shamil Shakirov for inspiring and stimulating discussions.
 All authors are supported in part
 by Vetenskapsr\r{a}det under grant \#2014-5517, by the STINT grant, and by the grant ''Geometry and Physics'' from the Knut and Alice Wallenberg foundation.
 The work of A.P. was supported in part by RFBR grants 16-01-00291, 18-31-20046 mol$\_$a$\_$ved and 19-01-00680 A.
}

\appendix

\section{Schur polynomials and partitions}\label{sec:schur}

Let us begin by introducing the notation for integer partitions. Consider the partition $\gamma=\{\gamma_1, \dots , \gamma_\bullet\}$ where $\gamma_1 \geq \dots \geq \gamma_\bullet > 0$ are positive integer numbers. Then $|\gamma|$ is the degree of the partition $\gamma$
\begin{equation}
 |\gamma| := \sum_{a\in\gamma}a,
\end{equation}
$l(\gamma)$ is the length of the partition
\begin{equation}
 l(\gamma) := \sum_{a\in\gamma}1,
\end{equation}
$\#_\gamma j$ is the number of parts $j$ in the partition $\gamma$
\begin{equation}
 \#_\gamma j := \sum_{a\in\gamma} \delta_{a,j},
\end{equation}
and $|\mathrm{Aut}(\gamma)|$ is the order of the automorphism group of the partition
\begin{equation}\label{eq:partition_notation}
 |\mathrm{Aut}(\gamma)| := \prod_{j=\gamma_\bullet}^{\gamma_1}(\#_\gamma j)!~.
\end{equation}
An alternative but equivalent description of such integer partitions is given in terms of \textit{Young diagrams} which are represented as left-aligned rows of boxes whose number is non-increasing (from top to bottom). The number of boxes on the $i$-th row is equal to the part $\gamma_i$ in the corresponding partition $\gamma$. Then the number of rows corresponds to the length of the partition and the total number of boxes to the degree.
As an example, consider the partition $\gamma = \{5,3,1,1\}$. This corresponds to the Young diagram
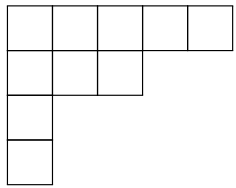
\begin{center}
\ydiagram{5,3,1,1}
\end{center}
which has degree $|\gamma|= 10$, length $l(\gamma) = 4$. The automorphism group of $\gamma$ corresponds to the group of permutations of all the rows with an equal number of boxes, and in this case it has order $|\mathrm{Aut}(\gamma)| = 2$.

We now wish to recall some properties of the multivariate Schur polynomials $s_{\gamma}(\{p_k\})$, which are labeled by partitions $\gamma$ and form a basis of the space of symmetric polynomials.
Firstly, they satisfy the Cauchy identity \cite[Chapter I, (4.3)]{Macdonald}
\begin{equation}
\begin{split}\label{eq:cauchy}
 \exp \left( \sum_{k=1}^\infty \frac{\tau_k p_k}{k} \right) = \sum_\gamma s_\gamma\left( \{ \tau_k\}\right) s_\gamma\left( \left\{p_k \right\} \right)~,
\end{split}
\end{equation}
where on the right hand side we sum over all partitions $\gamma$ and the variables $p_k$ are usually identified with traces of powers of some $N\times N$ matrices $\Phi$
\begin{equation}
 p_k=\mathrm{Tr}\;\Phi^k.
\end{equation}
In the main part of the paper we use the identification
\begin{equation}
 \Phi_{\alpha}=\mathrm{diag}(\lambda_{1,\alpha},\dots,\lambda_{N,\alpha})~,
\end{equation}
where $\alpha=1,2$ labels the two sets of variables \eqref{eq:exponentiated_variables} of the matrix model.

In the particular case when one adopts the plethystic substitution $\tau_k=z^k$, then on the right hand side of \eqref{eq:cauchy} the summation collapses to the symmetric partitions $\gamma=\{m\}$ only, i.e. those of length 1. Using the formula for symmetric Schur polynomials
\begin{equation} \label{eq:schur}
 s_{\{m\}}(p_1,\dots,p_m) =
 \sum_{\{\gamma\;\text{s.t.}\, |\gamma| = m \}}
 \frac{1}{|\mathrm{Aut}(\gamma)|}
 \prod_{i=1}^{l(\gamma)} \frac{p_i}{i}~,
\end{equation}
we have $s_{\{m\}}(\{\tau_k = z^k\})=z^m$, so that Cauchy's identity becomes
\begin{equation}\label{eq:symm-schur-poly}
 \exp\left(\sum_{k=1}^\infty \frac{z^k p_k}{k}\right) =
 \sum_{m=0}^\infty z^m s_{\{m\}}(p_1,\dots,p_m)~.
\end{equation}
More generally, Schur polynomials for arbitrary partitions $\gamma$ can be expressed using determinants as
\begin{equation}
 s_{\gamma}(\{\lambda_{i,\alpha}\}) = \frac{\det\limits_{i j} \, \lambda_{i,\alpha}^{N+\gamma_j-j}}{\det\limits_{i j} \, \lambda_{i,\alpha}^{N-j}}~,
\end{equation}
where now $\gamma_1\geq \dots \geq \gamma_N \geq 0$.

The first few Schur polynomials are given by 
\ytableausetup
{mathmode, boxsize=0.4em}
\begin{equation}
\begin{split}
s_{\emptyset} \left( \{p_k\} \right) & = s_{\{\}} \left( \{p_k\} \right) = \, 1 \\
s_{\ydiagram{1}} \left( \{p_k\} \right) & = s_{\{1\}} \left( \{p_k\} \right) = \, p_1 \\
s_{\ydiagram{2}} \left( \{p_k\} \right) & = s_{\{2\}} \left( \{p_k\} \right) = \, \frac{p_1^2+p_2}{2} \\
s_{\ydiagram{1,1}} \left( \{p_k\} \right) & = s_{\{1,1\}} \left( \{p_k\} \right) = \, \frac{p_1^2-p_2}{2} \\
s_{\ydiagram{3}} \left( \{p_k\} \right) & = s_{\{3\}} \left( \{p_k\} \right) = \, \frac{p_1^3 + 3p_1p_2 + 2p_3}{6} \\
s_{\ydiagram{2,1}} \left( \{p_k\} \right) & = s_{\{2,1\}} \left( \{p_k\} \right) = \, \frac{p_1^3 -p_3}{3} \\
s_{\ydiagram{1,1,1}} \left( \{p_k\} \right) & = s_{\{1,1,1\}} \left( \{p_k\} \right) = \, \frac{p_1^3 - 3p_1p_2 + 2p_3}{6} ~.
\end{split}
\end{equation}
We end this section by mentioning that integer partitions and Young diagrams have a deep relation with the representation theory of the symmetric group $S_n$ and, more importantly, the representation theory of $U(N)$ (and $SU(N)$). There is a 1-to-1 correspondence between partitions and irreducible representations\footnote{The highest weights $\gamma$ of the irreducibles of $U(N)$ are those that are integral ($\gamma_i\in\mathbb{Z}$) and dominant ($\gamma_1\geq\gamma_2\geq\dots\geq\gamma_N$).} of the unitary group and because of this it follows that Schur polynomials are exactly the irreducible characters of $U(N)$. For a given irreducible representation $\mathcal{R}_\gamma$ associated to the partition $\gamma$ and a group element $\Phi$, we have
\begin{equation}
 \mathrm{ch}_{\mathcal{R}_\gamma}(\Phi) = \mathrm{Tr}_{\mathcal{R}_\gamma}(\Phi) = s_\gamma(\{p_k=\mathrm{Tr}\;\Phi^k\}).
\end{equation}

\section{Special functions}

Here we collect a few well known facts about the special functions that we employ throughout the paper.
For additional details on the multiple sine functions we refer the reader to \cite{MR2010282}.

We first introduce the $q$-Pochhammer symbol, or $q$-shifted factorial, which is defined as
\begin{equation} \label{qpochhammer}
(x;q)_\infty = \mathe^{-\sum_{k>0}\frac{x^k}{k(1-q^k)}}=\prod_{k=0}^\infty (1-x q^k)
\end{equation}
with $x\in\mathbb{C}$ and $|q|<1$. The analytic continuation to the region $|q|>1$ is given by the formula 
\begin{equation}\label{eq:qpochhammer_continued}
(x;q)_\infty = \frac{1}{\left( q^{-1}x;q^{-1}\right)_\infty}~.
\end{equation}

Secondly we introduce the double sine function.
For $\underline{\omega}\equiv(\omega_1,\omega_2)\in\mathbb{C}^2$ with $\mathrm{Re}(\omega_\alpha)>0$ and $z\in\mathbb{C}$,
the double sine function is defined by the regularized product
\begin{equation}\label{eq:doubleSine}
  S_2 \left(z|\underline{\omega}\right) = \prod_{n_1, n_2
  \geq 0} \frac{n_1 \omega_1 + n_2 \omega_2 + z}{n_1 \omega_1 + n_2 \omega_2 + \omega
  - z}~,
\end{equation}
where $\omega = \omega_1 + \omega_2$. Let $\Lambda$ be the semi-lattice
\begin{equation}
 \Lambda = \omega_1\mathbb{Z}_{\geq 0} + \omega_2\mathbb{Z}_{\geq 0}~,
\end{equation}
then the double sine is a meromorphic function with zeroes and poles at
\begin{equation}\label{eq:PolesOfDoublesine}
\begin{aligned}
\text{zeroes}:\quad z \,&=\, -\Lambda\,,\\
\text{poles}:\quad z \,&=\, \omega+\Lambda\,~,
\end{aligned}
\end{equation}
as shown in Figure~\ref{fi:zerosAndPolesofdSine}.

\begin{figure}[h!]
  \begin{center}
    \begin{tikzpicture}[x=.7cm,y=.7cm]
    
    \def\a{2.0}
    \def\b{1.0}
    \def\c{0.5}
    \def\d{1.0}

    \draw [->, gray](-6.5,0)--(6.5,0) node [right, black] {$\mathbb{R}$};
    \draw [->, gray](0,-6.5)--(0,6.5) node [right, black] {$\mathrm{i}\mathbb{R}$};

    \draw [->, gray, dashed](-3*\a,-3*\b)--(3*\a,3*\b);
    \draw [->, gray, dashed](-6*\c,-6*\d)--(6*\c,6*\d);
    \draw [->, black](0,0)--(\a,\b);
    \draw [->, black](0,0)--(\c,\d);

    \node at (\a+0.5,\b-0.25) {$\omega_1$};
    \node at (\c-0.15,\d+0.5) {$\omega_2$};

    \draw [thin, gray](6,6)--(6,5) -- (7,5);
    \node at (6.65,5.6) {$z$};

    \node at (1*\a+1*\c,1*\b+1*\d) {$\times$};
    \node at (1*\a+2*\c,1*\b+2*\d) {$\times$};
    \node at (1*\a+3*\c,1*\b+3*\d) {$\times$};
    \node at (1*\a+4*\c,1*\b+4*\d) {$\times$};
    \node at (2*\a+1*\c,2*\b+1*\d) {$\times$};
    \node at (2*\a+2*\c,2*\b+2*\d) {$\times$};

    \fill[black] (-0*\a-0*\c,-0*\b-0*\d) circle(.15);
    \fill[white] (-0*\a-0*\c,-0*\b-0*\d) circle(.1);

    \fill[black] (-0*\a-1*\c,-0*\b-1*\d) circle(.15);
    \fill[white] (-0*\a-1*\c,-0*\b-1*\d) circle(.1);

    \fill[black] (-0*\a-2*\c,-0*\b-2*\d) circle(.15);
    \fill[white] (-0*\a-2*\c,-0*\b-2*\d) circle(.1);

    \fill[black] (-0*\a-3*\c,-0*\b-3*\d) circle(.15);
    \fill[white] (-0*\a-3*\c,-0*\b-3*\d) circle(.1);

    \fill[black] (-0*\a-4*\c,-0*\b-4*\d) circle(.15);
    \fill[white] (-0*\a-4*\c,-0*\b-4*\d) circle(.1);

    \fill[black] (-0*\a-5*\c,-0*\b-5*\d) circle(.15);
    \fill[white] (-0*\a-5*\c,-0*\b-5*\d) circle(.1);

    \fill[black] (-1*\a-0*\c,-1*\b-0*\d) circle(.15);
    \fill[white] (-1*\a-0*\c,-1*\b-0*\d) circle(.1);

    \fill[black] (-1*\a-1*\c,-1*\b-1*\d) circle(.15);
    \fill[white] (-1*\a-1*\c,-1*\b-1*\d) circle(.1);

    \fill[black] (-1*\a-2*\c,-1*\b-2*\d) circle(.15);
    \fill[white] (-1*\a-2*\c,-1*\b-2*\d) circle(.1);

    \fill[black] (-1*\a-3*\c,-1*\b-3*\d) circle(.15);
    \fill[white] (-1*\a-3*\c,-1*\b-3*\d) circle(.1);

    \fill[black] (-1*\a-4*\c,-1*\b-4*\d) circle(.15);
    \fill[white] (-1*\a-4*\c,-1*\b-4*\d) circle(.1);

    \fill[black] (-2*\a-0*\c,-2*\b-0*\d) circle(.15);
    \fill[white] (-2*\a-0*\c,-2*\b-0*\d) circle(.1);

    \fill[black] (-2*\a-1*\c,-2*\b-1*\d) circle(.15);
    \fill[white] (-2*\a-1*\c,-2*\b-1*\d) circle(.1);

    \fill[black] (-2*\a-2*\c,-2*\b-2*\d) circle(.15);
    \fill[white] (-2*\a-2*\c,-2*\b-2*\d) circle(.1);

    \end{tikzpicture}
    \caption{Zeroes $(\circ)$ and poles $(\times)$ of $S_2(z|\underline{\omega})$.
             All zeroes and poles are simple for generic $\omega_1,\omega_2$.
             }\label{fi:zerosAndPolesofdSine}
  \end{center}
\end{figure}
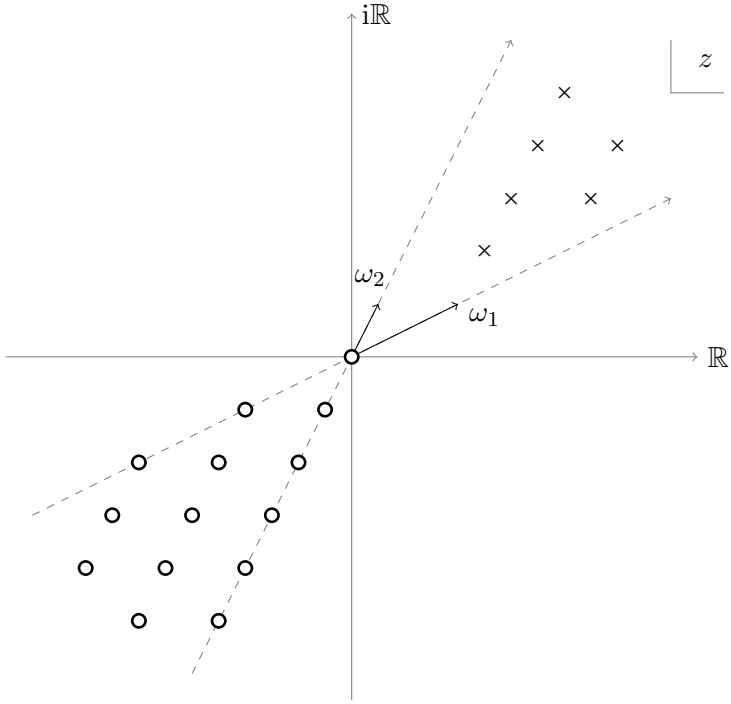
It can be shown that it satisfies the reflection identity
\begin{equation}
S_2 \left(z|\underline{\omega}\right)S_2 \left( \omega -z | \underline{\omega} \right) = 1~,
\end{equation}
as well as the first order finite difference equations
\begin{equation}\label{eq:quasiperiodicity}
\begin{split}
 S_2\left(z+\omega_1|\underline{\omega}\right) &=
 \frac{1}{2}\sin\left(\frac{\pi z}{\omega_{2}}\right)^{-1} S_2\left(z|\underline{\omega}\right) \\
 S_2\left(z+\omega_2|\underline{\omega}\right) &=
 \frac{1}{2}\sin\left(\frac{\pi z}{\omega_{1}}\right)^{-1} S_2\left(z|\underline{\omega}\right) ~.
\end{split}
\end{equation}
Moreover the double sine function is related to the hyperbolic gamma \cite{VdB} function $\Gamma_h$ by
\begin{equation}
 S_2(z|\underline{\omega}) = \Gamma_h(z;\omega_1,\omega_2)^{-1}~.
\end{equation}

Additionally, for $\mathrm{Im}(\omega_1/\omega_2)\neq 0$ one can express the double sine function as a product of two $q$-shifted factorials \cite{2003math......6164N}
\begin{equation} \label{eq:S2factorization}
 S_2 \left(z|\underline{\omega} \right) =
 \mathe^{\frac{\pi\mathi}{2}B_{22}\left( z | \underline{\omega} \right)}
 \left( \mathe^{\frac{2\pi \mathi }{\omega_1}z};\mathe^{\frac{2\pi \mathi }{\omega_1}\omega}\right)_\infty \left( \mathe^{\frac{2\pi \mathi }{\omega_2}z};\mathe^{\frac{2\pi \mathi }{\omega_2}\omega}\right)_\infty~,
\end{equation}
where $B_{22}\left(z|\underline{\omega}\right)$ is the double Bernoulli polynomial of degree 2
\begin{equation}
 B_{22}\left(z|\underline{\omega}\right) = \frac{1}{\omega_1\omega_2}
 \left(\left(z-\frac{\omega_1+\omega_2}{2}\right)^2-\frac{\omega_1^2+\omega_2^2}{12}\right)~.
\end{equation}

\section{Difference operator and shift of countour} \label{sec:contour_shift}

Let us consider the integral in \eqref{eq:generating_function}.
If we assume that the squashing parameters $\omega_1,\omega_2$ are both real and positive, then we can fix the integration contour $C$ to be a middle dimensional imaginary contour inside of $\mathbb{C}^N$. More precisely let $(\mathi\mathbb{R})^N$ be the subspace defined as
\begin{equation}
 (\mathi\mathbb{R})^N \assign \left\{ \underline{X} \in \mathbb{C}^N\,|\,\mathrm{Re}(X_i)=0,\,\mathrm{for}\, i=1,\dots,N \right\} \subset \mathbb{C}^N~,
\end{equation}
then the integration contour is prescribed by supersymmetric localization to be $C=(\mathi\mathbb{R})^N$.
With this choice the integrand falls off fast enough at complex infinity in the right-half-plane and one can compute the integral using the residue theorem by closing $C$ to the right. If there are poles lying exactly on the imaginary axis (of one of the variables) we give them a small real part or equivalently we deform slightly the contour to their left. 

For $n\geq -1$, we want to show the following identity between integrals:
\begin{equation}
\label{eq:shiftidentity}
\begin{aligned}
 \int_{C} \mathd X^N \, \shift \left[ \lambda_{i,\alpha}^{n} G_{i,\alpha}(\underline{\lambda})
 J(\underline{X}|\tautau)
 \right]
 & = \int_{\shift^{-1}C} \mathd X^N \, \left[ \lambda_{i,\alpha}^{n} G_{i,\alpha}(\underline{\lambda})
 J(\underline{X}|\tautau)
 \right] \\
 & = \int_{C} \mathd X^N \, \left[ \lambda_{i,\alpha}^{n} G_{i,\alpha}(\underline{\lambda})
 J(\underline{X}|\tautau)
 \right]~,
\end{aligned}
\end{equation}
where we consider one index $i$ and one order of $z$ at a time in \eqref{eq:starting_eq} (using \eqref{eq:z-series}). The non-trivial part of the identity is the second step where the shifted contour $\shift^{-1}C$ is deformed to the original contour $C$. The result of the integration will then be unchanged only if there are no poles in between the two contours. We now show that this is in fact true.

In order to show that there are no poles between $C$ and $\hat{\mathsf{M}}^{-1}_{i,\alpha}C$, we first notice that the shift operator $\shift$ only acts on the variable $\lambda_{i,\alpha}$ while leaving all other $\lambda_{j,\alpha}$ for $j \neq i$ unchanged. This means that we can simplify the problem by just focusing on the integration in the complex plane of $X_i$. Having fixed the index $i$ we can regard the other variables as background imaginary parameters. For definiteness we also take $\alpha=1$.

The poles that we should worry about are those coming from the function $G_{i,1}(\underline{\lambda})$, the measure $\Delta_S(\underline{X})$ and those coming from the antichiral multiplets.
The latter are the poles of $S_2(-X_i+\mu|\underline{\omega})^{-1}$ and are situated at the positions
\begin{equation}
 X_i = \mu + \Lambda~,
\end{equation}
which we assume to be to the right of the integration contour.\footnote{This is true if $\mathrm{Re}(\mu)>0$. If the real part of the mass is not positive then there will be a finite number poles to the left of the imaginary axes. In this case we can just modify the contour so that its endpoints are fixed but it goes to the left of these poles.}
After the shift, the poles of the function $\hat{\mathsf{M}}_{i,1} S_2(-X_i+\mu|\underline{\omega})^{-1}$ are situated at
\begin{equation}
 X_i = \omega_{2} + \mu + \Lambda~,
\end{equation}
so that clearly they are all still to the right of the contour $C$. A similar analysis follows for the mass $\nu$.

Next we look at the Vandermonde measure $\Delta_S(\underline{X})$.
Observe that the term $\prod_{j\neq i} S_2(X_i-X_j|\underline{\omega}) S_2(X_j-X_i|\underline{\omega})$ in the numerator only has zeros and no poles because of the identity
\begin{equation}
 S_2(z|\underline{\omega}) S_2(-z|\underline{\omega}) = - 4 \sin\left(\frac{\pi z}{\omega_1}\right) \sin\left(\frac{\pi z}{\omega_2}\right)~.
\end{equation}
The zeroes at $X_i = X_j + \omega_1\mathbb{Z}$ then cancel exactly the poles of the function $G_{i,1}(\underline{\lambda})$ coming from the denominator in \eqref{eq:G_function}.

The term $\prod_{j\neq i} S_2(X_i-X_j+M_\mathrm{a}|\underline{\omega})S_2(X_j-X_i+M_\mathrm{a}|\underline{\omega})$ in the denominator of $\Delta_S(\underline{X})$ is a bit more involved.
When one of the $X_j$ is infinite then the part of the Vandermonde which depends on $X_i$ and $X_j$ goes to $1$ and there are no poles because the adjoint mass becomes negligible and the numerator cancels with the denominator (in this regime the function $G_{i,1}(\underline{\lambda})$ is subject to the same kind of cancellation between poles and zeroes).
When the $X_j$ are finite we can regard them as finite imaginary shifts in the poles
\begin{equation}
 X_i = X_j \pm M_{\mathrm{a}} \pm \Lambda~,
\end{equation}
where for simplicity we assume that the adjoint mass $M_\mathrm{a}$ has a small and positive real part so that the contour separates the poles at $X_j+M_\mathrm{a}+\Lambda$ (on the right) from those at $X_j-M_\mathrm{a}-\Lambda$ (on the left).
The insertion of the function $G_{i,1}(\underline{\lambda})$ has the effect of removing the poles situated at the boundary of the semi-lattice
$X_j - M_{\mathrm{a}} - \Lambda$. If we now apply the shift operator $\hat{\mathsf{M}}_{i,1}$ we get that the shifted positions of the poles are:
\begin{equation}
 X_i = \omega_{2} + X_j \pm M_{\mathrm{a}} \pm \Lambda ~.
\end{equation}
We observe that those poles that would have crossed the imaginary axis because of the shift are precisely those at $X_i = X_j - M_\mathrm{a} - \omega_2\mathbb{Z}_{\geq0}$, i.e. those that are canceled by the numerator of $G_{i,1}(\underline{\lambda})$, so that the contributions to the residue computation are unchanged by application of $\hat{\mathsf{M}}_{i,1}$. In conclusion we have that the contour $\hat{\mathsf{M}}_{i,1}^{-1}C$ can be deformed to $C$ without crossing any poles and therefore the identity \eqref{eq:shiftidentity} holds.

\providecommand{\href}[2]{#2}\begingroup\raggedright\endgroup

\end{document}